\documentclass[conference]{IEEEtran}

\usepackage{cite}
\usepackage{amsmath,amssymb,amsfonts}
\usepackage{algorithmic}
\usepackage{graphicx}
\usepackage{textcomp}
\usepackage[dvipsnames]{xcolor}
\def\BibTeX{{\rm B\kern-.05em{\sc i\kern-.025em b}\kern-.08em
    T\kern-.1667em\lower.7ex\hbox{E}\kern-.125emX}}

\usepackage[symbol]{footmisc}
\usepackage{dirtytalk}
\usepackage{multirow}
\usepackage{float}
\usepackage{longtable}
\usepackage{hyperref}
\usepackage{tcolorbox}
\usepackage{forest}
\usepackage[utf8]{inputenc}
\usepackage{fontawesome}
\usepackage{pdfpages}
\usepackage{balance}

\AtBeginEnvironment{tcolorbox}{\small}
    
\begin{document}

\title{SoK: Privacy Risks and Mitigations in Retrieval-Augmented Generation Systems}

\author{\IEEEauthorblockN{Andreea-Elena Bodea\IEEEauthorrefmark{1}\IEEEauthorrefmark{2},
Stephen Meisenbacher\IEEEauthorrefmark{1}\IEEEauthorrefmark{3}, Alexandra Klymenko\IEEEauthorrefmark{4}, and
Florian Matthes\IEEEauthorrefmark{5}}
\IEEEauthorblockA{Technical University of Munich\\
School of Computation, Information and Technology\\
Garching, Germany\\
Email: \IEEEauthorrefmark{2}andreea.bodea@tum.de,
\IEEEauthorrefmark{3}stephen.meisenbacher@tum.de,
\IEEEauthorrefmark{4}alexandra.klymenko@tum.de,
\IEEEauthorrefmark{5}matthes@tum.de}}

\maketitle

\begin{abstract}
The continued promise of Large Language Models (LLMs), particularly in their natural language understanding and generation capabilities, has driven a rapidly increasing interest in identifying and developing LLM use cases. In an effort to complement the ingrained \say{knowledge} of LLMs, Retrieval-Augmented Generation (RAG) techniques have become widely popular. At its core, RAG involves the coupling of LLMs with domain-specific knowledge bases, whereby the generation of a response to a user question is augmented with contextual and up-to-date information. The proliferation of RAG has sparked concerns about data privacy, particularly with the inherent risks that arise when leveraging databases with potentially sensitive information. Numerous recent works have explored various aspects of privacy risks in RAG systems, from adversarial attacks to proposed mitigations. With the goal of surveying and unifying these works, we ask one simple question: \textit{What are the privacy risks in RAG, and how can they be measured and mitigated?} To answer this question, we conduct a systematic literature review of RAG works addressing privacy, and we systematize our findings into a comprehensive set of privacy risks, mitigation techniques, and evaluation strategies. We supplement these findings with two primary artifacts: a Taxonomy of RAG Privacy Risks and a RAG Privacy Process Diagram. Our work contributes to the study of privacy in RAG not only by conducting the first systematization of risks and mitigations, but also by uncovering important considerations when mitigating privacy risks in RAG systems and assessing the current maturity of proposed mitigations.
\end{abstract}


\section{Introduction}
\renewcommand*{\thefootnote}{\IEEEauthorrefmark{1}}
\footnotetext{These authors contributed equally.}
\renewcommand*{\thefootnote}{\arabic{footnote}}

With the seemingly ubiquitous recent advances in the areas of Artificial Intelligence and Natural Language Processing, predominantly spearheaded by modern Large Language Models (LLMs), the number of innovative use cases leveraging LLMs has grown at a likewise unfathomable rate \cite{liu2023summary,thirunavukarasu2023large,KASNECI2023102274}. LLMs have been pushed far beyond chatbots and translation tools, with impressive heights being reached in reasoning abilities, coding, multimodal generation, and agentic tasks. With this plethora of promising use cases, one persistent challenge with using LLMs is the inherent fact that these models are static and must inevitably be restricted by some \say{knowledge cutoff} \cite{li2024knowledge}, i.e., the most recent point in time at which the data used to train a model was collected, as well as other technical knowledge boundaries such as context size.

As a direct answer to this important problem, the Retrieval-Augmented Generation (RAG) paradigm \cite{lewis2020retrieval} has starkly risen in prevalence due to its simple yet effective method for incorporating external knowledge into generation with LLMs. By coupling such knowledge bases, which might contain domain-specific information or otherwise previously unseen data, the ability of LLMs to utilize information in context can be effectively leveraged to create query-response systems for answering user questions in an up-to-date and informed manner \cite{fan_survey_2024,zhao_retrieval-augmented_2024}. This plug-and-play method empowers users with knowledge bases to unlock the information contained within, and to make this knowledge accessible to others.

With this paradigm of interacting with LLMs, however, new risks are introduced when coupling private information with LLMs, and by extension, the RAG systems built around them. Concerns of data privacy arise when considering the direct interfacing of LLMs with potentially sensitive data contained within the connected databases, particularly in light of known LLM privacy issues \cite{wang_unique_2024}. Such risks, if exploited by malicious users, may result in the exposure of private information or the incorrect functioning of the RAG system, thus undermining the demonstrated promise of RAG \cite{deng_pandora_2024,zhou_trustworthiness_2024,fan_survey_2024}.

Many recent works have acknowledged the privacy risks in RAG \cite{li_matching_2024, zeng_good_2024}. While some works focus on exploring potential privacy risks, others propose specific methods for risk mitigation. Despite these existing works, however, there remains a lack of systematization of privacy risks in RAG, and moreover, of how these risks can be measured and mitigated. In this, we see it as crucial for researchers and practitioners alike to have a unified overview of privacy in RAG, and the lack of systematization despite numerous works in the field points to an important and timely research gap. 

We strive to understand the scope of privacy risks in RAG, uncovering the various ways in which recent works have measured privacy risks, particularly in the evaluation of proposed mitigation strategies. To gain such an overview, we conduct a Systematic Literature Review (SLR) of 72 recent papers at the intersection of RAG and privacy, studying the investigated risks, mitigations, and evaluation strategies of these works. We systematize this collection of 72 papers (Table \ref{tab:system}) and assess current mitigations (Table \ref{tab:mapping}). The findings of this survey lead to the creation of a Taxonomy of RAG Privacy Risks (Figure \ref{fig:taxonomy}), which enumerates risks and maps them to potential mitigations, and a RAG Privacy Process Diagram (Figure \ref{fig:process}), which illustrates a dynamic view of where along the RAG pipeline risks materialize and can be mitigated. 

Our literature survey teaches us that while many of the privacy risks associated with RAG can be considered under the umbrella of \textit{information leakage} and \textit{attacks}, these take many forms along the RAG pipeline. As such, the numerous types of proposed mitigation strategies can each be mapped to specific risks and, accordingly, to specific steps in the RAG process. Current mitigation efforts, however, have received varying degrees of attention, and we quantify both the \textit{relevance} and \textit{maturity} of privacy risk mitigations for RAG (Table \ref{tab:mapping}), showing a disparity in \textit{proposed} mitigations versus what may be considered \textit{mature} mitigations.

We contribute to the study of privacy in RAG as follows:
\begin{enumerate}
    \item We conduct the first systematic study to survey known RAG privacy risks, mitigations, and evaluation techniques as proposed by the recent literature.
    \item We offer two primary artifacts that systematize our main findings from the literature review: a Taxonomy of RAG Privacy Risks and a RAG Privacy Process Diagram.
    \item We also provide a comprehensive mapping of proposed mitigations to RAG privacy risks, and we quantify these mitigations according to their \textit{relevance} and \textit{maturity}.
    \item We publish a public repository of the surveyed papers, grey literature sources, and complete literature analysis: \url{https://github.com/sebischair/SoK-RAG-Privacy}.
\end{enumerate}

\section{Foundations}
RAG \cite{lewis2020retrieval} is an advanced framework designed to enhance the capabilities of LLMs by integrating external knowledge into the generation process. These systems address some of the inherent limitations of LLMs, such as hallucination, outdated information, and limited domain specificity \cite{li2024enhancing}. 

\begin{figure}[t]
    \centering
    \includegraphics[width=0.45\textwidth]{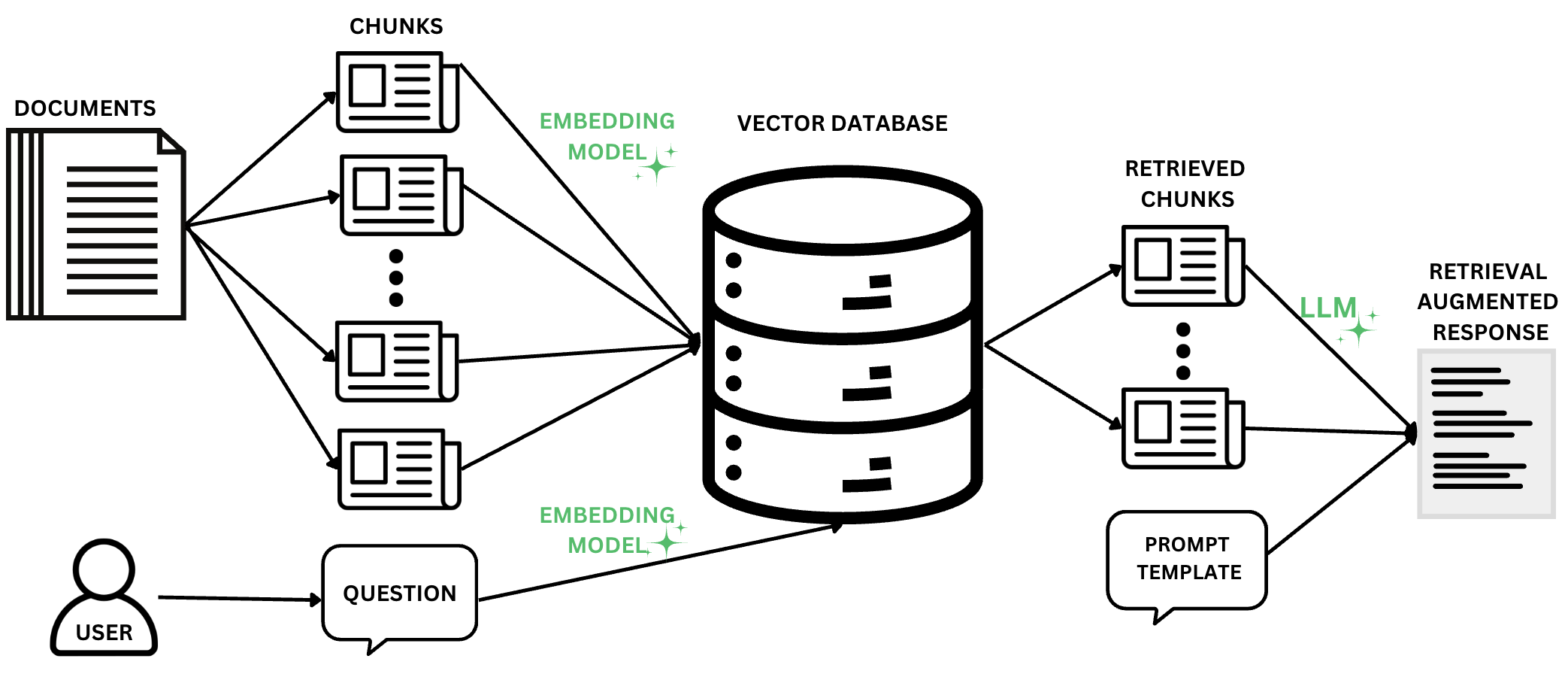}
    \caption{A typical Retrieval-Augmented Generation system.}
    \label{fig:RAG}
\end{figure}

As displayed in Figure \ref{fig:RAG}, the RAG process is characterized by three distinct stages: indexing, retrieval, and generation. During the indexing stage, raw data found in internal documents or external sources from the internet are cleaned, segmented into chunks, and converted into vector representations using embedding models. These vectors are then stored in a database that has been optimized for conducting similarity searches.
In the retrieval stage, the system receives a raw text user query, encodes it into a vector, and searches for the most semantically similar top-$k$ relevant text chunks in the vector database.  Finally, during the generation stage, the retrieved chunks are inserted into the LLM together with a pre-defined prompt. The LLM then produces a response by leveraging both its pre-trained knowledge and the additional retrieved context.

The RAG paradigm has been well-received by both research and practice, finding extensive use in a wide variety of domains and scenarios \cite{zhao-etal-2023-retrieving,zhao_retrieval-augmented_2024}. This includes, but is not limited to, dialogue systems, translation, and summarization, as well as coding and drug discovery. In addition to the \say{naive} RAG pipeline \cite{gao2024retrievalaugmentedgenerationlargelanguage}, as depicted in Figure \ref{fig:RAG}, there have been numerous proposals for enhancements in various areas along the pipeline \cite{gao2024retrievalaugmentedgenerationlargelanguage, zhao_retrieval-augmented_2024}, including advanced retrieval techniques and optimized hyperparameter selection.

As the RAG framework is not a single model but instead a confluence of techniques, the study of RAG includes a diverse variety of topics, which introduces complexities to implementing real-world systems. Particularly with the requirement to attach an external dataset for better response contextualization, new challenges may arise, both technical in nature (e.g., ensuring data quality) as well as increased concerns of privacy (i.e., protecting sensitive data within the provided database). Motivated by privacy aspects of RAG, and the lack of systematization on what exactly this entails, we seek to further the study of RAG with a comprehensive overview of RAG privacy risks and mitigations.

First, however, we establish an operational definition of \textit{privacy}, as this becomes important to systematizing and analyzing \textit{privacy risks} in RAG systems. 

\begin{tcolorbox}
    \textbf{Scope}: We ground our study in two important notions: \textit{private} and \textit{confidential} information, both of which play a role in the context of RAG. Private and confidential information differ in scope, context, and legal implications, though they are often used interchangeably. \textit{Private} information refers to the personal details of an individual or entity that are not intended for public dissemination but are meant to remain private. An example would be a person's medical history. \textit{Confidential} information also refers to sensitive data, but that shared between parties under an explicit or implicit agreement to be kept secret. An example would be propriety company information on internal projects. It is important to note that information can be private without being confidential, and vice versa; nevertheless, in this work, we view both private \textit{and} confidential information under the umbrella of \say{privacy}, or \say{data protection} (we do not distinguish between the two). Accordingly, in the context of our survey, we define privacy to mean \textit{the safeguarding of private information from improper disclosure and adversarial threats}.
\end{tcolorbox}

\section{Methodology}
In order to survey the current research on privacy in RAG, we conduct an SLR following the framework established by Kitchenham et al. \cite{10.5555/2994449}. The SLR enables the systematic investigation of a body of work, and as the first step to guide our investigation, we define one overarching research question:

\begin{enumerate}
    \item[\textbf{RQ}.] \textit{What are the privacy risks in RAG systems, and what techniques have been proposed to mitigate them?}
\end{enumerate}

Thus, we seek not only to discover risks and mitigations, but also to explore a mapping between the two and to identify strategies to measure the effectiveness of mitigations.

In line with the SLR methodology, we plan and conduct the review according to the following steps:

\begin{enumerate}
    \item \textbf{Selecting Databases for Primary Sources}: Both white and grey literature \cite{garousi2019guidelines} were utilized to ensure comprehensive coverage of this nascent topic. White literature sources included Google Scholar, ACM Digital Library, and IEEE Xplore. Grey literature was collected from Google Search and YouTube. This combination of academic and non-academic sources ensured a balanced perspective, capturing theoretical and practical insights.
    \item \textbf{Defining the Search Strings}: We developed two search strings to maximize relevant results while focusing the scope of our search. In particular, both strings contain two primary parts: terms referring to RAG and terms related to privacy or adversarial attacks, following our defined research scope. This resulted in the following:
        \begin{enumerate}
            \item[\textbf{S1}:] (``rag'' OR ``retrieval augmented'' OR ``augmented generation'') AND (``private'' OR "privacy'')
            \item[\textbf{S2}:] (``rag'' OR ``retrieval augmented'' OR ``augmented generation'') AND (``attack'')
        \end{enumerate}
    \item \textbf{Conducting the Search}: To maintain a relevant scope for the eventual selected papers, we limited the search results from Google Scholar to the first 150 results, and Google Search and YouTube to the first 50 results (five pages). Literature databases were limited to sources published from 2020 onward, in order to include only works after the formal introduction of RAG \cite{lewis2020retrieval}. For searching ACM DL and IEEE Xplore, the search strings were applied to the title and abstract only. The final search was performed in July 2025. Preprints were included only if they had existed for more than a year.
    \item \textbf{Exclusion Criteria}: Given the merged set of search results, exclusion criteria were applied via paper screening to achieve the final literature set. Beyond cursory quality checks and keeping only accessible sources, exclusion criteria involved removing duplicate studies and filtering out articles that do not explicitly address privacy-related issues in RAG systems. This was especially important for Google Scholar, since it does not allow for abstract searching. Therefore, a pre-filtering was performed on all retrieved results to remove sources that were clearly out of scope. The threshold of 150 was chosen after a pre-screen that revealed nearly all results thereafter to be clearly not relevant. Examples of irrelevant papers retrieved were papers implementing RAG systems and mentioning that privacy should be a concern without elaboration, i.e., where RAG systems are proposed for specific use cases for which privacy would be very important, but without any further justification or experimentation of privacy implications. Other examples include using RAG systems for LLM unlearning or to create attack graphs. Following these filtering steps, a final collection of \textbf{145} white literature sources was retained, as depicted in Table \ref{table:white_literature}. In addition, 48 Google Search results and 6 YouTube videos were deemed to be relevant; the complete list of grey literature sources is provided in the supplemental material in our repository.
    \item \textbf{Relevance Check}: Following exclusion, a more in-depth relevance check was conducted with the 145 selected sources. First, the abstract of each source was read to determine immediate relevance. If this was not clear, the full text of the work was screened, focusing on important aspects such as motivation, experiments, and discussion. If privacy was only covered tangentially, the source was filtered out. Specifically, the relevance check steps were:
    \begin{enumerate}
        \itemsep 0em
        \item The title of the paper was read. If relevance was clear, it was included. Otherwise,
        \item The abstract of the paper was read. If relevance became clear, it was included. Otherwise, 
        \item The paper full-text was screened.  If relevance became clear, it was included. Otherwise, it was deemed irrelevant.
    \end{enumerate}
    After this process, the final set of relevant papers consisted of \textbf{72} sources. We note that the entire process of filtering and relevance checking was done manually without the assistance of automatic tools.
    \item \textbf{Data Extraction}: Data extraction was primarily carried out by the lead researcher. Screening each primary source was prefaced by a reading of the abstract for familiarization with the work. Then, keywords such as `privacy' and `attack' were searched for, in order to find relevant points in the work for understanding the authors' perspective on privacy in RAG. Key information was extracted in a structured manner, including explicitly mentioned privacy risks, proposed mitigations, and experimental setup and evaluation. The included grey literature sources served to augment the findings from the white literature, often providing more accessible explanations of risks and attacks. These uncovered insights and other themes were discussed weekly with the complete research team over the course of the literature review process. A link to the complete structured data extraction results can be found in our public repository.
    \item \textbf{Synthesizing Results}: Each paper was annotated for \textit{privacy relevance} (1-3, with 3 being the most relevant), privacy focus (data leakage or adversarial manipulation, introduced next), and primary purpose (privacy attacks or mitigations). The findings from the literature review were also analyzed for three major categories following data extraction: (1) RAG privacy risks and attacks, (2) mitigation strategies for these privacy risks, and (3) evaluation datasets, tasks, and metrics for measuring privacy in RAG systems. This synthesis provided a comprehensive understanding of the current research landscape and highlighted gaps for further investigation. 
\end{enumerate}

\begin{table}[t]
\centering
\caption{An overview of the collected primary sources from the SLR, both before and after applying filtering and exclusion criteria, as well as following the final relevance check.}
\small
\resizebox{0.75\linewidth}{!}{
\begin{tabular}{ |c|c|c|c|c| } 
 \hline
& \multicolumn{2}{c|}{\textbf{S1}} & \multicolumn{2}{c|}{\textbf{S2}} \\ \cline{2-5} 
& Before & After & Before & After \\ \hline 
 Google Scholar & 150 & 43 & 150 & 72 \\ 
 ACM DL & 16 & 9 & 1 & 1 \\ 
 IEEE Xplore & 61 & 20 & 10 & 0 \\  \hline 
 Before exclusion & \multicolumn{4}{c|}{388} \\  \hline
 After exclusion & \multicolumn{4}{c|}{145} \\  \hline
 Final relevant set & \multicolumn{4}{c|}{72} \\ 
 \hline
\end{tabular}
}
\label{table:white_literature}
\end{table}

We systematize in Table \ref{tab:system} the structured data extracted from the 72 papers. These insights underpin our findings, detailed next, which include two main artifacts, the Taxonomy of RAG Privacy Risks and the RAG Privacy Process Diagram. 

\begin{figure*}[t!]
    \centering 
    \resizebox{0.9\textwidth}{!}{
    \begin{forest} for tree={%
            draw=gray, minimum width=1cm, minimum height=.5cm, rounded corners=3,
            text height=1.5ex, text depth=0ex,
            grow=east,reversed,
            edge={gray},
            parent anchor=east,
            child anchor=west,
            if n children=0{tier=last}{}
        }
    [\textbf{RAG Privacy Risks},draw=black 
        [Leakage,draw=RoyalBlue 
            [Data Leakage\cite{wang_unique_2024, zhou_trustworthiness_2024, zeng_good_2024,cohen_unleashing_2024, xian_understanding_2024, cho_typos_2024, yu_textual_2024,vizgirda_socialgenpod_2024,ng_simplyretrieve_2023,chuang_retrieval_2024,stefano_rag_2024,hu_privacy-preserved_2024,huang2023privacy_implications,chaudhari_phantom_2024,wang_poisoned_2024,xian_vulnerability_2024,pasquini_neural_2024,li2023multi,zeng_mitigating_2024,wang_mememo_2024,liu_mask-based_2024,anderson_is_my_data_2024,zhang_human-imperceptible_2024,tan_glue_2024,li_generating_2024,qi2024follow,Ward_Harguess_2025,Kulshreshtha_Choudhary_Taneja_Verma_2025,Fang_Qiao_Shi_An_2025,Chen_Tackman_Setälä_Poranen_Zhang_2025,Yu_Liu_Denny_Bergen_Liut_2025, He_Tang_Zhang_Zhou_Su_2025,Hussain_2025,He_Liu_Hou_Jiang_Li_2025_PRESS,Grislain_2025,mehtasecure,Nandagopal_2025,golatkar_cpr_2024,xue_db-gpt_2024,dou_design_2024},draw=RoyalBlue  [Dataset Leakage, draw=RoyalBlue ][Vector Database Leakage, draw=RoyalBlue][Retrieved Chunk Leakage, draw=RoyalBlue][Answer Leakage, draw=RoyalBlue]][Prompt Leakage \cite{wang_unique_2024,stefano_rag_2024,Ward_Harguess_2025,Nandagopal_2025,zyskind_dont_2023,agarwal_prompt_leakage_2024}, draw=RoyalBlue]]
        [Adversarial Manipulation,draw=Maroon 
            [Data Poisoning Attacks \cite{wang_unique_2024,deng_pandora_2024,zhou_trustworthiness_2024,cohen_unleashing_2024,xian_understanding_2024,cho_typos_2024,chaudhari_phantom_2024,wang_poisoned_2024,xian_vulnerability_2024,zhang_human-imperceptible_2024,tan_glue_2024,Ward_Harguess_2025,Kulshreshtha_Choudhary_Taneja_Verma_2025,mehtasecure,Nandagopal_2025,xue_badrag_2024,zhang_agent_2024,chen_agentpoison_2024,zhu_atm_2024,chen_black-box_2024,jiao_can_2024,xiang_certifiably_2024,roychowdhury_confusedpilot_2024,fang_enhancing_2024,altinisik_exploiting_2024,ju_flooding_2024,zhang_hijackrag_2024,shafran_machine_2024,zou_poisonedrag_2024,jiang_tc-ragturing-complete_2024,kuppa_manipulating_2024,nazary2025resource,Mo_Tang_Lin_Zhou_Li_2025,Nazary_Deldjoo_Noia_2025,Zhang_Zhang_Lou_Wu_Wang_Chen_2025,Jiao_Wang_Yang_2025,Zhang_Xin_Fang_Liu_Yi_Li_Liu_2025},draw=Maroon][Backdoor Attacks \cite{wang_unique_2024,zhou_trustworthiness_2024,chaudhari_phantom_2024,Nandagopal_2025,cheng_trojanrag_2024,xue_badrag_2024,clop_backdoored_2024},draw=Maroon][Data Extraction Attacks \cite{wang_unique_2024,zhou_trustworthiness_2024,zeng_good_2024,huang2023privacy_implications,qi2024follow,peng_data_2024},draw=Maroon][Jailbreak Attacks \cite{wang_unique_2024,deng_pandora_2024,zhou_trustworthiness_2024,wang_poisoned_2024,Nandagopal_2025,pfrommer_ranking_2024},draw=Maroon][LLM Extraction/Inversion Attacks \cite{wang_unique_2024,Ward_Harguess_2025},draw=Maroon][Prompt Extraction Attacks \cite{wang_unique_2024,agarwal_prompt_leakage_2024},draw=Maroon][Membership Inference Attacks \cite{wang_unique_2024,zhou_trustworthiness_2024,cohen_unleashing_2024,liu_mask-based_2024,anderson_is_my_data_2024,li_generating_2024,Ward_Harguess_2025,mehtasecure,Feng_Zhang_Tian_Xu_Zhang_Zhu_Ding_Liu_2025_RAGLeak},draw=Maroon][Prompt Injection Attacks \cite{wang_unique_2024,stefano_rag_2024,pasquini_neural_2024,qi2024follow,Ward_Harguess_2025,Nandagopal_2025,clop_backdoored_2024,shafran_machine_2024,kuppa_manipulating_2024,nazary2025resource,pfrommer_ranking_2024,bondarenko_llm_2024,hu_prompt_2024},draw=Maroon]]
    ]
    \end{forest}
    }
    \caption{The Taxonomy of RAG Privacy Risks. While we highlight the two-sided nature of privacy risks in RAG, leakage and adversarial manipulation, we focus specifically on \textit{data leakage} and its mitigation. Adversarial manipulation, or attacks, primarily serve as a means to realize threats posed by leakage.}
    \label{fig:taxonomy}
\end{figure*}
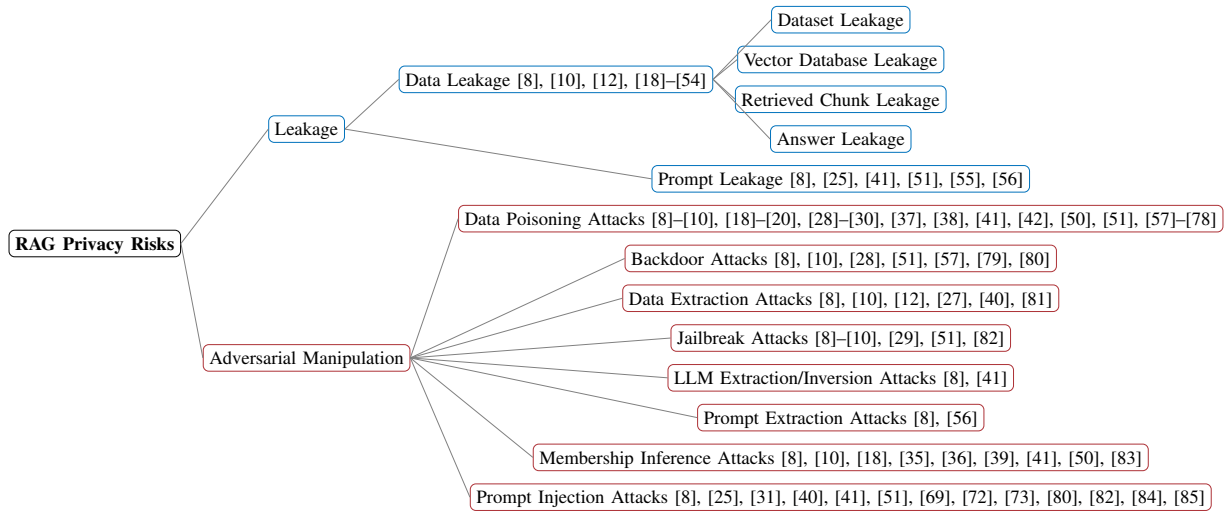

\section{RAG Privacy Risks and Mitigations}
We present the main findings from our survey of 72 literature sources, which take the form of privacy \textit{risks}, \textit{mitigations}, and \textit{evaluation strategies}. We illustrate that the privacy risks in RAG can be categorized into two major categories, \textit{leakage} and \textit{adversarial manipulation}. As much of the surveyed literature also focuses on mitigations, we provide a direct mapping of mitigation strategies to risk points along the RAG pipeline. Finally, we break down evaluation strategies for measuring privacy protection into two primary aspects: \textit{datasets} and \textit{metrics} used for evaluation. 

\subsection{A Taxonomy of RAG Privacy Risks}
The review of relevant literature addressing privacy in RAG systems revealed two predominant ways in which privacy risks can be perceived. The first refers to the idea of \textit{attacks}, or malicious attempts to disrupt, disable, or misuse RAG systems. We call these \textit{Adversarial Manipulation} risks. We include these in our presented taxonomy, but as these attack vectors exists on the boundary between security and privacy risks, we only briefly introduce them below. We refer the reader to the cited works (Figure \ref{fig:taxonomy}) for more details.

\begin{itemize}
    \item \textbf{Jailbreak Attacks} use specially designed prompts or sequences to bypass a RAG system’s safety filters, enabling the generation of harmful, toxic, or restricted content. These attacks exploit the generative model’s contextual sensitivity to subvert built-in content moderation policies.
    \item \textbf{Backdoor Attacks} introduce malicious triggers during training or fine-tuning, which remain dormant until activated by specific inputs. In RAG systems, these may persist even in retrieved document chunks.
    \item \textbf{Data Poisoning Attacks} involve corrupting training or retrieval datasets through the malicious injection of adversarial examples, mislabeled data, or misleading content. 
    \item \textbf{Prompt Injection Attacks} involve embedding adversarial content within prompts to manipulate system behavior. These attacks exploit the interpretative flexibility of generative models, potentially causing the system to execute unintended instructions or disclose sensitive information.
    \item \textbf{Membership Inference Attacks} aim to determine whether a specific data point was a part of a model's training set or, in the context of RAG, if such data is present in the knowledge base. The retrieval component of RAG can exacerbate this risk by exposing responses tied to unique data samples.
    \item In a \textbf{Data Extraction Attack}, adversaries exploit model outputs to reconstruct sensitive data. These attacks challenge the privacy of both the retrieval and generative components, especially in systems lacking robust access controls or output sanitization.
    \item \textbf{Prompt Extraction} entails the reconstruction of user prompts from system behavior or responses. Such attacks threaten to enable unauthorized parties to access or infer other users' inputs, which can contain private or confidential information.
    \item \textbf{LLM Extraction/Inversion Attacks} target the underlying parameters (i.e., knowledge representation) of language models. By systematically querying a RAG system, adversaries may infer embedded facts or even reconstruct portions of the training corpus.
\end{itemize}

\begin{tcolorbox}
    \textbf{Key finding}: A survey of privacy in RAG intersects with the study of \textit{adversarial manipulation}, which covers a variety of adversarial attacks that, in some form, may affect the privacy of the end user or compromise the confidentiality of data held by private entities. In the remainder of this work, we view adversarial manipulation in tandem with \textit{leakage} risks.
\end{tcolorbox}

The second aspect of privacy in RAG is complex, and it relates to the idea of \textit{leakage} resulting from RAG's inner workings. Leakage poses a privacy risk in the potential for exposure of sensitive information originating from the data stored within the RAG system, or conversely, data inputted into the system via user prompts that may be leaked at a later point. Beyond this, we noticed that while many works cover this topic nominally, few works describe explicitly what form of data leakage they are protecting against, as opposed to defined attack vectors as introduced above.

Due to the unique and sequential nature of the RAG pipeline, however, leakage can originate from many points, and we find that this origination point is directly tied to the point at which mitigation strategies would be applied. For example, mitigating \textit{dataset leakage} would imply that protections are implemented immediately at the database level, further implying that the repercussions of such protections are acceptable for the functioning of the ensuing pipeline. However, if personal information is required to retrieve relevant encoded chunks in the vector database, mitigation measures might be applied after this stage, thereby mitigating \textit{retrieved chunk leakage}. We therefore learn of the significant distinction between types of data leakage, which becomes important for defining risks and mapping appropriate mitigation strategies. 

The complete Taxonomy of RAG Privacy Risks is presented in Figure \ref{fig:taxonomy}. We introduce the five uncovered types of leakage risks in RAG systems, four of which relate to the potential risks from data being passed through the RAG system internally, and one relating to the risks of user-provided data via prompts. In this, we also introduce associated attack types, thus showing the interconnection between Leakage and Adversarial Manipulation. Finally, we include our findings on suggestions for mitigations as reported in the literature; a mapping of mitigations to risks is found in Section \ref{sec:mapping}. 

\subsubsection{Dataset leakage}
Dataset leakage is an issue particularly when proprietary or sensitive information is stored using unsafe storage solutions. Leakage can occur through external exposure, internal access control failures, or both.

One risk is accidental exposure of proprietary data due to insecure storage practices. If contributors store sensitive documents in unprotected cloud storage, shared drives, or even email attachments, unauthorized individuals may gain access. Unlike structured SQL databases with well-defined access controls, many traditional storage solutions rely on manual access management, increasing the likelihood of misconfigurations that lead to data breaches. Moreover, publicly available datasets, such as those scraped from the internet, may unknowingly contain private information, blurring the line between open-source and confidential data.

Another concern is inadequate internal access controls, which can lead to both intentional and accidental data exposure. In many organizations, employees in both technical and non-technical departments may have unrestricted access to all stored documents. This level of access poses multiple risks. First, employees might unintentionally modify metadata or tags, making critical documents unavailable or incorrectly prioritized during retrieval. Second, unrestricted editing rights could lead to the unintentional inclusion of sensitive data, potentially affecting downstream generated responses.

Non-malicious data leakage often occurs when actors inadvertently include PII or other sensitive information in documents without realizing these files will later be indexed into the RAG system. A major failure point is the insufficient removal or masking of personal data. If PII is not properly sanitized, confidential details such as names, addresses, phone numbers, or legal case specifics may become part of the system's retrieval process. This can lead to unauthorized exposure when an AI model retrieves and presents sensitive information in response to user queries.

\textbf{Mitigations.} Addressing dataset leakage in RAG systems requires a multi-layered approach. Organizations must implement robust access control policies \cite{dou_design_2024, vizgirda_socialgenpod_2024, cohen_unleashing_2024}, ensuring that only authorized personnel can view or edit sensitive documents (and their associated vector embeddings). This can be supplemented with distributed data storage solutions or specialized cloud architectures \cite{lewis2020retrieval,zyskind_dont_2023}. Automated PII detection, filtering, and redaction tools (i.e., \textit{anonymization}) should be integrated into the data ingestion pipeline to prevent accidental exposure \cite{huang2023privacy_implications, cohen_unleashing_2024, zhou_trustworthiness_2024}. Beyond PII handling, rewriting or rephrasing techniques can be used to modify the original documents while maintaining their semantic meaning \cite{chen_agentpoison_2024,xiang_certifiably_2024,tan_glue_2024, zhang_hijackrag_2024,zhang_human-imperceptible_2024,shafran_machine_2024,chaudhari_phantom_2024,zou_poisonedrag_2024}. Alternatively, synthetic data could be used in lieu of the original data, assuming this achieves acceptable performance \cite{zeng_mitigating_2024}. Furthermore, monitoring mechanisms can help identify and mitigate the risks associated with data poisoning and backdoor injection.

\begin{tcolorbox}
    \textbf{Key finding}: We learn of two primary aspects of dataset leakage in RAG systems, leaking data to RAG end users via PII or sensitive information, and exposing sensitive or confidential information to unauthorized internal users. While the majority of the literature proposing mitigations focuses on the former, such as through redaction or anonymization, there has been much less attention paid to the latter. Furthermore, no works investigate the interconnectedness of database leakage mitigation, for example, how proper anonymization can serve as a supporting tool to database access controls.
\end{tcolorbox}

\subsubsection{Vector database leakage}
The risk of vector database leakage stems from the cases when proprietary or sensitive data is stored in vector databases. Unlike traditional databases, vector storage enables powerful semantic search but also introduces new risks. One critical issue is embedding model memorization, where the model retains patterns from its training data. If the embedding model has been exposed to proprietary documents, attackers can probe the system with crafted queries to retrieve sensitive information. This risk increases when embeddings are not properly sanitized, potentially allowing unauthorized users to reconstruct proprietary data from the model's learned representations. 

A direct risk arises when sensitive documents are stored and then retrieved without proper controls. If a query closely matches confidential content, the system may return private information embedded in the vector store. For example, a request about financial agreements between companies could unintentionally reveal contract terms. Furthermore, attackers can refine their queries to bypass simple safeguards. 

An overlooked risk is misconfigured database access, where weak authentication or improper permission settings expose stored embeddings or document chunks to unauthorized users. Exploits thus can extract embeddings, reconstruct sensitive data, or query for proprietary documents. 

\textbf{Mitigations.} To protect embeddings, Differential Privacy (DP) techniques and synthetic data can help mitigate model memorization \cite{yu_textual_2024, zeng_mitigating_2024}. To prevent the improper access of sensitive information, the information can be redacted before indexing, access can be restricted based on user roles \cite{dou_design_2024, vizgirda_socialgenpod_2024, cohen_unleashing_2024}, and query filtering can help to block the retrieval of classified content. Other proposals include the injection of redundant non-sensitive examples into the vector stores, as well as simple duplication.

\begin{tcolorbox}
    \textbf{Key finding}: An important aspect of RAG systems is the transformation of texts into a vector database store. Privacy issues here are mainly rooted in the leakage of information stored in vector embeddings, despite the inherent belief that embeddings may successfully obfuscate data. While we find prior works focusing on access control and user roles for vector databases, we find relatively few works that propose mitigations at the embedding level, outside of exploratory works on DP and synthetic data. Importantly, we also find no mention of protecting the \textit{mapped} text in the vector databases, i.e., the original text chunks to which the vectors correspond. This calls for important future research that aims to balance semantic coherence and privacy protection in embedding representations and stored texts of RAG data.
\end{tcolorbox}

\subsubsection{Retrieved chunk leakage}
Retrieved chunk leakage occurs when private information is exposed in system responses due to the retrieval of sensitive or proprietary content. This issue arises when the retrieval process pulls confidential information from stored documents, incorporates such information in the response generation, and presents a potentially leaky answer to users.
One major risk is internal manipulation by actors with access to the retrieval pipeline. Technical stakeholders, if malicious, could manipulate retrieval processes to prioritize certain chunks, leading to biased or unauthorized exposure of confidential data. 

\textbf{Mitigations.} Mitigating retrieval-based leakage requires robust retrieval strategies \cite{cohen_unleashing_2024,zeng_mitigating_2024,zeng_good_2024,golatkar_cpr_2024} and distance metrics \cite{zeng_mitigating_2024, zeng_good_2024, zhou_trustworthiness_2024} to ensure that retrieved chunks are both relevant and safe for disclosure. This could be achieved by altering indexing mechanisms, modifying metadata, or improving the ranking step that determines which chunks are most relevant to a query. For example, Differential Privacy techniques can be applied at the cross-attention stage in reranking, adding controlled noise to reduce the likelihood of retrieving highly sensitive content \cite{liang_differential_2024}.

\begin{tcolorbox}
    \textbf{Key finding}: Another unique aspect to RAG systems is the retrieval stage, in which chunks of text data are retrieved, often via similarity of user prompts to stored embedding representations. Initial work has been performed looking mainly at mitigations in the employed \textit{retrieval strategy}. We find, however, that further research is warranted, especially as the retrieval stage serves both as the \say{last line of defense} before LLM answer generation and also as a potential privacy-utility \say{balance point}, where utility loss from early risk mitigation on the database level (e.g., via anonymization) can be avoided while still preventing unwanted leakage to the LLM and in the output to the user.
\end{tcolorbox}

\subsubsection{Answer leakage}
As the final stage in the RAG pipeline, answer leakage can occur when private or sensitive information is unintentionally revealed in the response generated by the system. Even if access to the retrieved chunks is restricted, the LLM may still incorporate confidential data into its output, leading to unintentional exposure.

One primary concern is the content of the generated answer itself. If an LLM receives sensitive chunks without proper filtering, it may produce responses that disclose confidential or private data, such as confidential discussions. An equally important risk is the storage of generated responses, particularly in logging systems, conversation histories, or cached outputs. If responses with sensitive data are stored, they can be retrieved in later queries or accessed by unauthorized users, further exacerbating privacy risks.

\textbf{Mitigations.} To mitigate answer leakage, organizations can consider local deployment of RAG models to ensure full control over data handling and prevent external exposure. Implementing response safeguards \cite{zhang_agent_2024} such as post-processing filters, fact-checking mechanisms \cite{ju_flooding_2024}, and structured validation can help detect and redact sensitive information before it is displayed. Additionally, enforcing source citation allows transparency, ensuring that sensitive responses are traced back to their origins, making it easier to flag and prevent private data from being included in outputs.

\begin{tcolorbox}
    \textbf{Key finding}: Investigating privacy risks at the answer generation stage in RAG systems is complex, in the way that data leakage can be propagated from retrieved chunks \textit{as well as} from ingrained sensitive data in LLMs that may be triggered by certain RAG inputs. Current proposed mitigations seem to be heavily geared towards the risk that the LLMs pose, where local LLM deployment and guardrails comprise the vast majority of methods. This, however, leaves a relatively wide gap in mitigating data leaked \textit{from the RAG system itself}, i.e., from the (vector) database and retrieved chunks. 
\end{tcolorbox}

\subsubsection{Prompt leakage}
Prompt leakage is another critical privacy concern in RAG systems, particularly when user prompts contain sensitive information.
This is especially pertinent when a RAG system logs or stores prompts for future reference. If queries contain private information, they may persist in conversation history, be cached for optimization, or even be retained in memory, potentially making them accessible in unintended contexts. This becomes especially problematic when responses based on those queries and retrieved chunks are also stored. A user within the same session may unintentionally retrieve sensitive information from previous exchanges, and another user, whether intentionally or not, could later trigger references to previously stored prompts or responses.

\textbf{Mitigations.} To enhance privacy, one can integrate mechanisms such as anonymization, PII removal, and query filtering to prevent sensitive data from persisting in stored queries \cite{huang2023privacy_implications, cohen_unleashing_2024, zhou_trustworthiness_2024}. Paraphrasing, or rewriting prompts before processing them can further reduce risks while maintaining query intent \cite{chen_agentpoison_2024, zhang_human-imperceptible_2024, zeng_mitigating_2024}. Additionally, privacy-aware models with augmented prompts can help models to recognize and redact sensitive input dynamically. Finally, prompts can be distributed using techniques such as Multi-Party Computation \cite{zyskind_dont_2023}, ensuring that no single server receives the entire prompt with potentially sensitive information.

\begin{tcolorbox}
    \textbf{Key finding}: While mitigating privacy risks from user inputs to RAG systems may appear to be the most straightforward of the uncovered leakage types, we observe in the literature that the focus primarily lies on scrutinizing the content of the prompt text itself. On the other hand, although the literature points out that the \textit{context} surrounding the user prompt is important, namely in determining which knowledge and information a particular user is privy to, there is a scarcity of mitigations that make these considerations. This also becomes a crucial factor in ensuring that storage logs of user prompts are privatized correctly.
\end{tcolorbox}

\begin{table}[htbp]
\centering
\caption{A systematization of the 72 reviewed papers. We classify each of the papers along several axes: year of publication, number of citations (as of September 24, 2025), privacy relevance (scale of 1-3, with 3 meaning directly relevant), privacy focus (leakage \faUmbrella \ or adversarial manipulation \faUserSecret), primary purpose of the paper (attack $\dagger$, mitigation \faShield, or neither ``-''), whether the paper introduces mitigations (implements \textcolor{Green}{\faCheck}, mentions \textcolor{Yellow}{\faComment}, or none \textcolor{Red}{\faRemove}), and whether experiments are run (yes \textcolor{Green}{\faCheck} or no \textcolor{Red}{\faRemove}), code is available (if yes, linked to \faCode), and discussions on practical applicability (latency, overhead, etc.) are included (yes \textcolor{Green}{\faCheck} or no \textcolor{Red}{\faRemove}).}
\label{tab:system}
\resizebox{0.96\linewidth}{!}{
\begin{tabular}{|l|c|c|c|c|c|c|c|c|c|}
\hline
\textbf{Key} & \textbf{Year} & \textbf{Cit.} & \textbf{Rel.} & \textbf{Focus} & \textbf{Purpose} & \textbf{Mit.?} & \textbf{Exp.?} & \textbf{Code?} & \textbf{Disc.?}  \\ \hline
\cite{Nazary_Deldjoo_Noia_2025} & 2025 & 29 & 2 & \faUserSecret & $\dagger$ & \textcolor{Green}{\faCheck} & \textcolor{Green}{\faCheck} & \href{https://github.com/atenanaz/Poison-RAG}{\faCode} & \textcolor{Red}{\faRemove} \\
\cite{liu_mask-based_2024} & 2025 & 13 & 3 & \faUmbrella + \faUserSecret & $\dagger$ & \textcolor{Green}{\faCheck} & \textcolor{Green}{\faCheck} &  & \textcolor{Red}{\faRemove} \\
\cite{Grislain_2025} & 2025 & 5 & 3 & \faUmbrella & \faShield & \textcolor{Green}{\faCheck} & \textcolor{Red}{\faRemove} & \href{https://github.com/sarus-tech/dp-rag}{\faCode} & \textcolor{Red}{\faRemove} \\
\cite{Jiao_Wang_Yang_2025} & 2025 & 4 & 2 & \faUserSecret & $\dagger$ & \textcolor{Red}{\faRemove} & \textcolor{Green}{\faCheck} &  & \textcolor{Red}{\faRemove} \\
\cite{Yu_Liu_Denny_Bergen_Liut_2025} & 2025 & 4 & 1 & \faUmbrella & - & \textcolor{Green}{\faCheck} & \textcolor{Red}{\faRemove} &  & \textcolor{Green}{\faCheck} \\
\cite{Zhang_Xin_Fang_Liu_Yi_Li_Liu_2025} & 2025 & 3 & 2 & \faUserSecret & \faShield & \textcolor{Green}{\faCheck} & \textcolor{Green}{\faCheck} &  & \textcolor{Red}{\faRemove} \\
\cite{Nandagopal_2025} & 2025 & 2 & 3 & \faUserSecret & \faShield & \textcolor{Yellow}{\faComment} & \textcolor{Red}{\faRemove} &  & \textcolor{Green}{\faCheck} \\
\cite{He_Tang_Zhang_Zhou_Su_2025} & 2025 & 1 & 3 & \faUmbrella & \faShield & \textcolor{Green}{\faCheck} & \textcolor{Green}{\faCheck} &  & \textcolor{Red}{\faRemove} \\
\cite{nazary2025resource} & 2025 & 1 & 3 & \faUserSecret & $\dagger$ & \textcolor{Yellow}{\faComment} & \textcolor{Green}{\faCheck} & \href{https://anonymous.4open.science/r/Poison-RAG-Plus-6B03/README.md}{\faCode} & \textcolor{Red}{\faRemove} \\
\cite{He_Liu_Hou_Jiang_Li_2025_PRESS} & 2025 & 1 & 3 & \faUmbrella & \faShield & \textcolor{Green}{\faCheck} & \textcolor{Green}{\faCheck} &  & \textcolor{Green}{\faCheck} \\
\cite{Kulshreshtha_Choudhary_Taneja_Verma_2025} & 2025 & 1 & 1 & \faUserSecret & - & \textcolor{Yellow}{\faComment} & \textcolor{Red}{\faRemove} &  & \textcolor{Red}{\faRemove} \\
\cite{Ward_Harguess_2025} & 2025 & 0 & 3 & \faUmbrella + \faUserSecret & \faShield & \textcolor{Yellow}{\faComment} & \textcolor{Red}{\faRemove} &  & \textcolor{Red}{\faRemove} \\
\cite{mehtasecure} & 2025 & 0 & 3 & \faUmbrella +   \faUserSecret & \faShield & \textcolor{Yellow}{\faComment} & \textcolor{Red}{\faRemove} &  & \textcolor{Red}{\faRemove} \\
\cite{Mo_Tang_Lin_Zhou_Li_2025} & 2025 & 0 & 2 & \faUserSecret & $\dagger$ & \textcolor{Green}{\faCheck} & \textcolor{Green}{\faCheck} &  & \textcolor{Red}{\faRemove} \\
\cite{Fang_Qiao_Shi_An_2025} & 2025 & 0 & 2 & \faUmbrella & \faShield & \textcolor{Green}{\faCheck} & \textcolor{Green}{\faCheck} &  & \textcolor{Green}{\faCheck} \\
\cite{Chen_Tackman_Setälä_Poranen_Zhang_2025} & 2025 & 0 & 2 & \faUmbrella & \faShield & \textcolor{Green}{\faCheck} & \textcolor{Green}{\faCheck} & \href{https://github.com/bingxiangch/thesis_auth_rag/}{\faCode} & \textcolor{Red}{\faRemove} \\
\cite{Feng_Zhang_Tian_Xu_Zhang_Zhu_Ding_Liu_2025_RAGLeak} & 2025 & 0 & 2 & \faUserSecret & $\dagger$ & \textcolor{Yellow}{\faComment} & \textcolor{Green}{\faCheck} &  & \textcolor{Red}{\faRemove} \\
\cite{Hussain_2025} & 2025 & 0 & 1 & \faUmbrella + \faUserSecret & \faShield & \textcolor{Green}{\faCheck} & \textcolor{Red}{\faRemove} &  & \textcolor{Red}{\faRemove} \\
\hline 
\cite{xiong_benchmarking_2024} & 2024 & 305 & 1 & \faUmbrella & - & \textcolor{Yellow}{\faComment} & \textcolor{Red}{\faRemove} & \href{https://github.com/Teddy-XiongGZ/MedRAGhttps://github.com/Teddy-XiongGZ/MedRAG}{\faCode} & \textcolor{Red}{\faRemove} \\
\cite{zou_poisonedrag_2024} & 2024 & 214 & 3 & \faUserSecret & $\dagger$ & \textcolor{Green}{\faCheck} & \textcolor{Green}{\faCheck} & \href{https://github.com/sleeepeer/PoisonedRAG}{\faCode} & \textcolor{Green}{\faCheck} \\
\cite{zeng_good_2024} & 2024 & 132 & 3 & \faUmbrella & $\dagger$ & \textcolor{Green}{\faCheck} & \textcolor{Green}{\faCheck} & \href{https://github.com/phycholosogy/RAGprivacy}{\faCode} & \textcolor{Red}{\faRemove} \\
\cite{chen_agentpoison_2024} & 2024 & 127 & 1 & \faUmbrella + \faUserSecret & $\dagger$ & \textcolor{Green}{\faCheck} & \textcolor{Green}{\faCheck} & \href{https://github.com/BillChan226/AgentPoison}{\faCode} & \textcolor{Red}{\faRemove} \\
\cite{golatkar_cpr_2024} & 2024 & 99 & 1 & \faUmbrella & \faShield & \textcolor{Green}{\faCheck} & \textcolor{Green}{\faCheck} &  & \textcolor{Green}{\faCheck} \\
\cite{xue_badrag_2024} & 2024 & 87 & 3 & \faUserSecret & $\dagger$ & \textcolor{Green}{\faCheck} & \textcolor{Green}{\faCheck} &  & \textcolor{Red}{\faRemove} \\
\cite{xiang_certifiably_2024} & 2024 & 76 & 2 & \faUmbrella +   \faUserSecret & \faShield & \textcolor{Green}{\faCheck} & \textcolor{Green}{\faCheck} &  & \textcolor{Green}{\faCheck} \\
\cite{chaudhari_phantom_2024} & 2024 & 74 & 3 & \faUmbrella + \faUserSecret & $\dagger$ & \textcolor{Yellow}{\faComment} & \textcolor{Green}{\faCheck} &  & \textcolor{Red}{\faRemove} \\
\cite{cheng_trojanrag_2024} & 2024 & 64 & 3 & \faUserSecret & $\dagger$ & \textcolor{Yellow}{\faComment} & \textcolor{Green}{\faCheck} & \href{https://github.com/Charles-ydd/TrojanRAG}{\faCode} & \textcolor{Red}{\faRemove} \\
\cite{zhang_agent_2024} & 2024 & 64 & 1 & \faUmbrella + \faUserSecret & $\dagger$ & \textcolor{Green}{\faCheck} & \textcolor{Green}{\faCheck} & \href{https://github.com/agiresearch/ASB}{\faCode} & \textcolor{Red}{\faRemove} \\
\cite{deng_pandora_2024} & 2024 & 62 & 3 & \faUmbrella +   \faUserSecret & $\dagger$ & \textcolor{Red}{\faRemove} & \textcolor{Green}{\faCheck} &  & \textcolor{Red}{\faRemove} \\
\cite{fang_enhancing_2024} & 2024 & 56 & 2 & \faUserSecret & \faShield & \textcolor{Green}{\faCheck} & \textcolor{Green}{\faCheck} & \href{https://github.com/calubkk/RAAT}{\faCode} & \textcolor{Red}{\faRemove} \\
\cite{cho_typos_2024} & 2024 & 50 & 3 & \faUmbrella & $\dagger$ & \textcolor{Green}{\faCheck} & \textcolor{Green}{\faCheck} & \href{https://github.com/zomss/GARAG}{\faCode} & \textcolor{Red}{\faRemove} \\
\cite{qi2024follow} & 2024 & 48 & 2 & \faUmbrella & $\dagger$ & \textcolor{Green}{\faCheck} & \textcolor{Green}{\faCheck} & \href{https://github.com/zhentingqi/rag-privacy}{\faCode} & \textcolor{Red}{\faRemove} \\
\cite{pasquini_neural_2024} & 2024 & 46 & 3 & \faUmbrella +   \faUserSecret & $\dagger$ & \textcolor{Yellow}{\faComment} & \textcolor{Green}{\faCheck} & \href{https://github.com/pasquinidario/LLM_NeuralExec}{\faCode} & \textcolor{Red}{\faRemove} \\
\cite{ju_flooding_2024} & 2024 & 46 & 2 & \faUmbrella + \faUserSecret & $\dagger$ & \textcolor{Yellow}{\faComment} & \textcolor{Green}{\faCheck} & \href{https://github.com/Jometeorie/KnowledgeSpread}{\faCode} & \textcolor{Red}{\faRemove} \\
\cite{xue_db-gpt_2024} & 2024 & 46 & 1 & \faUmbrella & - & \textcolor{Yellow}{\faComment} & \textcolor{Red}{\faRemove} & \href{https://github.com/eosphoros-ai/DB-GPT}{\faCode} & \textcolor{Green}{\faCheck} \\
\cite{anderson_is_my_data_2024} & 2024 & 41 & 2 & \faUmbrella & $\dagger$ & \textcolor{Green}{\faCheck} & \textcolor{Green}{\faCheck} &  & \textcolor{Red}{\faRemove} \\
\cite{hu_prompt_2024} & 2024 & 38 & 2 & \faUserSecret & $\dagger$ & \textcolor{Green}{\faCheck} & \textcolor{Green}{\faCheck} & \href{https://github.com/Hadise-zb/Prompt-Perturbation-in-Retrieval-AugmentedGeneration/tree/main}{\faCode} & \textcolor{Green}{\faCheck} \\
\cite{zhang_human-imperceptible_2024} & 2024 & 37 & 2 & \faUmbrella + \faUserSecret & $\dagger$ & \textcolor{Yellow}{\faComment} & \textcolor{Green}{\faCheck} &  & \textcolor{Red}{\faRemove} \\
\cite{zeng_mitigating_2024} & 2024 & 29 & 3 & \faUmbrella & \faShield & \textcolor{Green}{\faCheck} & \textcolor{Green}{\faCheck} & \href{https://anonymous.4open.science/r/RAG-SAGE-E175/Readme.md}{\faCode} & \textcolor{Red}{\faRemove} \\
\cite{liang_differential_2024} & 2024 & 29 & 1 & \faUmbrella & \faShield & \textcolor{Green}{\faCheck} & \textcolor{Red}{\faRemove} &  & \textcolor{Red}{\faRemove} \\
\cite{jiang_tc-ragturing-complete_2024} & 2024 & 26 & 1 & \faUserSecret & $\dagger$ & \textcolor{Green}{\faCheck} & \textcolor{Green}{\faCheck} & \href{https://https://github.com/Artessay/SAMA.git}{\faCode} & \textcolor{Red}{\faRemove} \\
\cite{liu_can_2024} & 2024 & 25 & 1 & \faUmbrella & - & \textcolor{Yellow}{\faComment} & \textcolor{Red}{\faRemove} &  & \textcolor{Green}{\faCheck} \\
\cite{tan_glue_2024} & 2024 & 23 & 2 & \faUmbrella +   \faUserSecret & $\dagger$ & \textcolor{Green}{\faCheck} & \textcolor{Green}{\faCheck} &  & \textcolor{Red}{\faRemove} \\
\cite{li_generating_2024} & 2024 & 17 & 2 & \faUmbrella + \faUserSecret & $\dagger$ & \textcolor{Green}{\faCheck} & \textcolor{Green}{\faCheck} &  & \textcolor{Red}{\faRemove} \\
\cite{jiao_can_2024} & 2024 & 17 & 1 & \faUserSecret & $\dagger$ & \textcolor{Green}{\faCheck} & \textcolor{Green}{\faCheck} &  & \textcolor{Red}{\faRemove} \\
\cite{stefano_rag_2024} & 2024 & 16 & 3 & \faUmbrella + \faUserSecret & $\dagger$ & \textcolor{Green}{\faCheck} & \textcolor{Green}{\faCheck} &  & \textcolor{Red}{\faRemove} \\
\cite{pfrommer_ranking_2024} & 2024 & 16 & 3 & \faUserSecret & $\dagger$ & \textcolor{Red}{\faRemove} & \textcolor{Green}{\faCheck} &  & \textcolor{Green}{\faCheck} \\
\cite{wang_mememo_2024} & 2024 & 16 & 1 & \faUmbrella & \faShield & \textcolor{Green}{\faCheck} & \textcolor{Red}{\faRemove} & \href{https://github.com/spfrommer/cse-ranking-manipulationhttps://github.com/spfrommer/ragdoll-data-pipeline}{\faCode} & \textcolor{Red}{\faRemove} \\
\cite{shafran_machine_2024} & 2024 & 15 & 3 & \faUserSecret & $\dagger$ & \textcolor{Green}{\faCheck} & \textcolor{Green}{\faCheck} & \href{https://poloclub.github.io/mememo}{\faCode} & \textcolor{Green}{\faCheck} \\
\cite{cohen_unleashing_2024} & 2024 & 15 & 3 & \faUmbrella + \faUserSecret & $\dagger$ & \textcolor{Yellow}{\faComment} & \textcolor{Green}{\faCheck} &  & \textcolor{Red}{\faRemove} \\
\cite{zhu_atm_2024} & 2024 & 15 & 2 & \faUmbrella +   \faUserSecret & \faShield & \textcolor{Green}{\faCheck} & \textcolor{Green}{\faCheck} & \href{https://github.com/StavC/UnleashingWorms-ExtractingData}{\faCode} & \textcolor{Green}{\faCheck} \\
\cite{chen_black-box_2024} & 2024 & 15 & 2 & \faUserSecret & $\dagger$ & \textcolor{Yellow}{\faComment} & \textcolor{Green}{\faCheck} & \href{https://github.com/chuhac/ATM-RAG}{\faCode} & \textcolor{Red}{\faRemove} \\
\cite{wang_poisoned_2024} & 2024 & 15 & 2 & \faUmbrella +   \faUserSecret & $\dagger$ & \textcolor{Red}{\faRemove} & \textcolor{Green}{\faCheck} &  & \textcolor{Red}{\faRemove} \\
\cite{peng_data_2024} & 2024 & 11 & 3 & \faUmbrella & $\dagger$ & \textcolor{Green}{\faCheck} & \textcolor{Green}{\faCheck} & \href{https://github.com/CAM-FSS/jailbreak-langchain}{\faCode} & \textcolor{Red}{\faRemove} \\
\cite{roychowdhury_confusedpilot_2024} & 2024 & 11 & 2 & \faUserSecret & $\dagger$ & \textcolor{Green}{\faCheck} & \textcolor{Green}{\faCheck} &  & \textcolor{Green}{\faCheck} \\
\cite{clop_backdoored_2024} & 2024 & 10 & 2 & \faUserSecret & $\dagger$ & \textcolor{Red}{\faRemove} & \textcolor{Green}{\faCheck} &  & \textcolor{Red}{\faRemove} \\
\cite{zhang_hijackrag_2024} & 2024 & 9 & 2 & \faUserSecret & $\dagger$ & \textcolor{Green}{\faCheck} & \textcolor{Green}{\faCheck} &  & \textcolor{Red}{\faRemove} \\
\cite{agarwal_prompt_leakage_2024} & 2024 & 8 & 3 & \faUmbrella & \faShield & \textcolor{Green}{\faCheck} & \textcolor{Green}{\faCheck} &  & \textcolor{Green}{\faCheck} \\
\cite{xian_vulnerability_2024} & 2024 & 6 & 3 & \faUmbrella +   \faUserSecret & $\dagger$ & \textcolor{Green}{\faCheck} & \textcolor{Green}{\faCheck} &  & \textcolor{Red}{\faRemove} \\
\cite{yu_textual_2024} & 2024 & 6 & 3 & \faUmbrella & \faShield & \textcolor{Green}{\faCheck} & \textcolor{Green}{\faCheck} &  & \textcolor{Red}{\faRemove} \\
\cite{bondarenko_llm_2024} & 2024 & 5 & 2 & \faUserSecret & $\dagger$ & \textcolor{Red}{\faRemove} & \textcolor{Green}{\faCheck} &  & \textcolor{Red}{\faRemove} \\
\cite{chuang_retrieval_2024} & 2024 & 5 & 1 & \faUmbrella & \faShield & \textcolor{Green}{\faCheck} & \textcolor{Red}{\faRemove} & \href{https://github.com/alebondarenko/llm-robustness}{\faCode} & \textcolor{Red}{\faRemove} \\
\cite{vizgirda_socialgenpod_2024} & 2024 & 5 & 1 & \faUmbrella & \faShield & \textcolor{Green}{\faCheck} & \textcolor{Red}{\faRemove} &  & \textcolor{Red}{\faRemove} \\
\cite{hu_privacy-preserved_2024} & 2024 & 4 & 2 & \faUmbrella & \faShield & \textcolor{Green}{\faCheck} & \textcolor{Green}{\faCheck} & \href{https://github.com/Vidminas/socialgenpod/}{\faCode} & \textcolor{Red}{\faRemove} \\
\cite{shan_certifying_2024} & 2024 & 2 & 2 & \faUmbrella & \faShield & \textcolor{Red}{\faRemove} & \textcolor{Red}{\faRemove} & \href{https://github.com/HKUST-KnowComp/PrivateNGDB}{\faCode} & \textcolor{Red}{\faRemove} \\
\cite{kuppa_manipulating_2024} & 2024 & 2 & 2 & \faUserSecret & $\dagger$ & \textcolor{Red}{\faRemove} & \textcolor{Green}{\faCheck} &  & \textcolor{Red}{\faRemove} \\
\cite{xian_understanding_2024} & 2024 & 2 & 2 & \faUmbrella & $\dagger$ & \textcolor{Green}{\faCheck} & \textcolor{Green}{\faCheck} &  & \textcolor{Red}{\faRemove} \\
\cite{altinisik_exploiting_2024} & 2024 & 2 & 1 & \faUmbrella + \faUserSecret & \faShield & \textcolor{Green}{\faCheck} & \textcolor{Green}{\faCheck} &  & \textcolor{Green}{\faCheck} \\
\cite{dou_design_2024} & 2024 & 1 & 1 & \faUmbrella & - & \textcolor{Yellow}{\faComment} & \textcolor{Red}{\faRemove} &  & \textcolor{Red}{\faRemove} \\
\hline
\cite{li2023multi} & 2023 & 456 & 2 & \faUmbrella & $\dagger$ & \textcolor{Yellow}{\faComment} & \textcolor{Green}{\faCheck} &  & \textcolor{Red}{\faRemove} \\
\cite{huang2023privacy_implications} & 2023 & 56 & 3 & \faUmbrella & \faShield & \textcolor{Green}{\faCheck} & \textcolor{Green}{\faCheck} & \href{https://github.com/HKUST-KnowComp/LLM-Multistep-Jailbreak}{\faCode} & \textcolor{Red}{\faRemove} \\
\cite{zyskind_dont_2023} & 2023 & 14 & 1 & \faUmbrella & \faShield & \textcolor{Green}{\faCheck} & \textcolor{Green}{\faCheck} & \href{https://github.com/Princeton-SysML/kNNLM_privacy}{\faCode} & \textcolor{Green}{\faCheck} \\
\hline
\end{tabular}
}
\end{table}

\begin{table}[t!]
\centering
\caption{A mapping of proposed mitigation strategies from the literature to the privacy risks associated with different stages of the RAG pipeline, as well as with our survey adversarial manipulation threats. \textit{Rel.} and \textit{Mat.} denote the relevance and maturity scores, which are normalized to between 0 and 1 to show relative ranking. For ease of interpretation, we also color-code the scores into four bins: [0, 0.25): \textcolor{Red}{\faCircle}, [0.25, 0.50): \textcolor{Orange}{\faCircle}, [0.50, 0.75): \textcolor{Yellow}{\faCircle}, and [0.75, 1.00]: \textcolor{Green}{\faCircle}.}
\resizebox{0.48\textwidth}{!}{
\footnotesize
\begin{tabular}{|c|p{0.95\linewidth}|c|c|}
\hline
\textbf{Privacy Risk} & \multicolumn{1}{|c|}{\textbf{Proposed Mitigation Solutions}} & \textbf{Rel.} & \textbf{Mat.} \\ \hline
\multicolumn{4}{|c|}{\textbf{Leakage}} \\ \hline
\multirow{7}{*}{\textbf{Dataset}} & Anonymization \cite{zhou_trustworthiness_2024,cohen_unleashing_2024,huang2023privacy_implications, li2023multi,Ward_Harguess_2025,Kulshreshtha_Choudhary_Taneja_Verma_2025,Fang_Qiao_Shi_An_2025,Hussain_2025,mehtasecure,xue_db-gpt_2024} & \textcolor{Green}{\faCircle} 0.77 & \textcolor{Red}{\faCircle} 0.23 \\
 & Synthetic Data \cite{zeng_mitigating_2024} & \textcolor{Red}{\faCircle} 0.12 & \textcolor{Orange}{\faCircle} 0.67\\
 & Text Rewriting / Rephrasing \cite{zeng_mitigating_2024, zhang_human-imperceptible_2024, chen_agentpoison_2024} & \textcolor{Red}{\faCircle} 0.19 & \textcolor{Red}{\faCircle} 0.22\\
 & Text Summarization \cite{zhou_trustworthiness_2024,zeng_good_2024,zeng_mitigating_2024} & \textcolor{Red}{\faCircle} 0.19 & \textcolor{Red}{\faCircle} 0.22\\
 & Differential Privacy \cite{yu_textual_2024,Kulshreshtha_Choudhary_Taneja_Verma_2025,Fang_Qiao_Shi_An_2025,He_Tang_Zhang_Zhou_Su_2025,Grislain_2025} & \textcolor{Yellow}{\faCircle} 0.50 & \textcolor{Orange}{\faCircle} 0.27\\ 
 & Distributed Data Storage / Hybrid Cloud Architectures \cite{lewis2020retrieval,zyskind_dont_2023} & \textcolor{Red}{\faCircle} 0.15 & \textcolor{Red}{\faCircle} 0.00\\
 & Encryption \cite{Fang_Qiao_Shi_An_2025,mehtasecure,Nandagopal_2025} & \textcolor{Red}{\faCircle} 0.19 & \textcolor{Red}{\faCircle} 0.22\\ 
 \hline
\multirow{3}{*}{\textbf{Vector Database}}  & Access Control \cite{cohen_unleashing_2024,vizgirda_socialgenpod_2024,Fang_Qiao_Shi_An_2025,Chen_Tackman_Setälä_Poranen_Zhang_2025,Hussain_2025,mehtasecure,Nandagopal_2025,dou_design_2024} & \textcolor{Yellow}{\faCircle} 0.65 & \textcolor{Red}{\faCircle} 0.17\\
 & Redundant Benign Knowledge Base / Duplication \cite{stefano_rag_2024,huang2023privacy_implications,jiao_can_2024} & \textcolor{Orange}{\faCircle} 0.42 & \textcolor{Red}{\faCircle} 0.22 \\
 & Synthetic Data \cite{zeng_mitigating_2024} & \textcolor{Red}{\faCircle} 0.12 & \textcolor{Yellow}{\faCircle} 0.67\\ \hline
\multirow{3}{*}{\textbf{Retrieved Chunks}} & Retrieval Strategy (e.g., \# of chunks, re-ranking) \cite{zeng_good_2024,cohen_unleashing_2024,zeng_mitigating_2024,golatkar_cpr_2024} & \textcolor{Orange}{\faCircle} 0.38 & \textcolor{Orange}{\faCircle} 0.33\\
 & Distance Metric Strategy (e.g., thresholds) \cite{zhou_trustworthiness_2024, zeng_good_2024, zeng_mitigating_2024} & \textcolor{Red}{\faCircle} 0.19 & \textcolor{Red}{\faCircle} 0.22\\
 & Differential Privacy (in re-ranking) \cite{liang_differential_2024} & \textcolor{Red}{\faCircle} 0.12 & \textcolor{Red}{\faCircle} 0.00 \\ \hline
\multirow{3}{*}{\textbf{Answer}} & Local LLM Deployment \cite{wang_mememo_2024,Fang_Qiao_Shi_An_2025,Chen_Tackman_Setälä_Poranen_Zhang_2025,Yu_Liu_Denny_Bergen_Liut_2025, xue_db-gpt_2024, dou_design_2024, xiong_benchmarking_2024, liu_can_2024} & \textcolor{Yellow}{\faCircle} 0.58 & \textcolor{Red}{\faCircle} 0.13\\
 & Safeguards / Guardrails \cite{Yu_Liu_Denny_Bergen_Liut_2025,Nandagopal_2025,xue_db-gpt_2024,zhang_agent_2024} & \textcolor{Orange}{\faCircle} 0.27 & \textcolor{Orange}{\faCircle} 0.42\\
 & Fact-checking / Source Citation \cite{ju_flooding_2024} & \textcolor{Red}{\faCircle} 0.10 & \textcolor{Red}{\faCircle} 0.00\\ \hline
\multirow{4}{*}{\textbf{Prompt}} & Anonymization (Remove/Mask/Filter PII) \cite{zhou_trustworthiness_2024,cohen_unleashing_2024, huang2023privacy_implications,Ward_Harguess_2025,mehtasecure} & \textcolor{Orange}{\faCircle} 0.31 & \textcolor{Red}{\faCircle} 0.13\\
 & Text Rewriting / Paraphrasing / Regrouping \cite{zeng_mitigating_2024, zhang_human-imperceptible_2024,qi2024follow,chen_agentpoison_2024} & \textcolor{Orange}{\faCircle} 0.35 & \textcolor{Orange}{\faCircle} 0.33 \\
 & Prompt Augmentation / Guardrails \cite{stefano_rag_2024} & \textcolor{Red}{\faCircle} 0.04 & \textcolor{Red}{\faCircle} 0.00\\
 & Multi-party Computation (MPC) \cite{zyskind_dont_2023} & \textcolor{Red}{\faCircle} 0.12 & \textcolor{Red}{\faCircle} 0.00\\ \hline
\multicolumn{4}{|c|}{\textbf{Adversarial Manipulation (Attacks)}} \\ \hline
& Anonymization \cite{zhou_trustworthiness_2024} & \textcolor{Red}{\faCircle} 0.00 & \textcolor{Red}{\faCircle} 0.00\\
\multirow{3}{*}{\textbf{Backdoor}} & Distance Metric Strategy \cite{zhou_trustworthiness_2024} & \textcolor{Red}{\faCircle} 0.00 & \textcolor{Red}{\faCircle} 0.00\\
& Text Summarization \cite{zhou_trustworthiness_2024} & \textcolor{Red}{\faCircle} 0.00 & \textcolor{Red}{\faCircle} 0.00\\ 
& Knowledge Expansion \cite{cheng_trojanrag_2024} & \textcolor{Red}{\faCircle} 0.00 & \textcolor{Red}{\faCircle} 0.00\\ 
& Detection of Anomaly Clusters \cite{cheng_trojanrag_2024} & \textcolor{Red}{\faCircle} 0.00 & \textcolor{Red}{\faCircle} 0.00\\ 
 \hline
\multirow{5}{*}{\textbf{Data Extraction}} & Anonymization \cite{zhou_trustworthiness_2024, huang2023privacy_implications} & \textcolor{Red}{\faCircle} 0.15 & \textcolor{Orange}{\faCircle} 0.33\\
 & Distance Metric Strategy \cite{zhou_trustworthiness_2024} & \textcolor{Red}{\faCircle} 0.00 & \textcolor{Red}{\faCircle} 0.00\\
 & Prompt Augmentation / Guardrails \cite{qi2024follow} & \textcolor{Red}{\faCircle} 0.12 & \textcolor{Yellow}{\faCircle} 0.67\\
 & Text Summarization \cite{zhou_trustworthiness_2024} & \textcolor{Red}{\faCircle} 0.00 & \textcolor{Red}{\faCircle} 0.00\\ 
 & Fine Tuning \cite{peng_data_2024} & \textcolor{Red}{\faCircle} 0.12 & \textcolor{Orange}{\faCircle} 0.33\\ \hline
 \multirow{13}{*}{\textbf{Data Poisoning}} 
 & Access Control \cite{chaudhari_phantom_2024,mehtasecure,Nandagopal_2025,roychowdhury_confusedpilot_2024} & \textcolor{Red}{\faCircle} 0.23 & \textcolor{Red}{\faCircle} 0.00 \\
 & Anonymization \cite{zhou_trustworthiness_2024,cohen_unleashing_2024,Ward_Harguess_2025,Kulshreshtha_Choudhary_Taneja_Verma_2025,mehtasecure} & \textcolor{Red}{\faCircle} 0.19 & \textcolor{Red}{\faCircle} 0.00 \\
 & Distance Metric Strategy
\cite{zhou_trustworthiness_2024,xian_vulnerability_2024,Kulshreshtha_Choudhary_Taneja_Verma_2025} & \textcolor{Red}{\faCircle} 0.19 & \textcolor{Red}{\faCircle} 0.11\\
& Fact-checking \cite{ju_flooding_2024} & \textcolor{Red}{\faCircle} 0.00 & \textcolor{Red}{\faCircle} 0.00\\
& Guardrails \cite{chaudhari_phantom_2024,Nandagopal_2025,zhang_agent_2024} & \textcolor{Red}{\faCircle} 0.19 & \textcolor{Red}{\faCircle} 0.22 \\
  & Retrieval Strategy \cite{cohen_unleashing_2024,zhang_human-imperceptible_2024,xiang_certifiably_2024,shafran_machine_2024,Zhang_Xin_Fang_Liu_Yi_Li_Liu_2025} & \textcolor{Orange}{\faCircle} 0.42 & \textcolor{Red}{\faCircle} 0.20  \\
 & Rewriting / Rephrasing \cite{chaudhari_phantom_2024,zhang_human-imperceptible_2024,tan_glue_2024,chen_agentpoison_2024,xiang_certifiably_2024, zhang_hijackrag_2024,shafran_machine_2024, zou_poisonedrag_2024,Zhang_Xin_Fang_Liu_Yi_Li_Liu_2025} & \textcolor{Green}{\faCircle} 1.00 & \textcolor{Orange}{\faCircle} 0.37\\
 & Text Summarization \cite{zhou_trustworthiness_2024,Nazary_Deldjoo_Noia_2025} & \textcolor{Red}{\faCircle} 0.15 & \textcolor{Red}{\faCircle} 0.17\\
& Text Duplication / Knowledge Expansion \cite{Mo_Tang_Lin_Zhou_Li_2025,Nazary_Deldjoo_Noia_2025,Zhang_Zhang_Lou_Wu_Wang_Chen_2025} & \textcolor{Orange}{\faCircle} 0.42 & \textcolor{Orange}{\faCircle} 0.33\\
& Text Filtering \cite{xian_understanding_2024,chaudhari_phantom_2024,tan_glue_2024,zou_poisonedrag_2024,Mo_Tang_Lin_Zhou_Li_2025,Zhang_Zhang_Lou_Wu_Wang_Chen_2025} & \textcolor{Green}{\faCircle} 0.77 & \textcolor{Orange}{\faCircle} 0.39\\
& Perplexity \cite{chaudhari_phantom_2024,xue_badrag_2024,chen_agentpoison_2024,shafran_machine_2024,zou_poisonedrag_2024,Mo_Tang_Lin_Zhou_Li_2025} & \textcolor{Green}{\faCircle} 0.85 & \textcolor{Orange}{\faCircle} 0.38\\
& Adversarial Training \cite{cohen_unleashing_2024,cho_typos_2024,Ward_Harguess_2025,zhu_atm_2024,fang_enhancing_2024,altinisik_exploiting_2024} & \textcolor{Yellow}{\faCircle} 0.62 & \textcolor{Orange}{\faCircle} 0.33\\ 
& Grammar Checker \cite{cho_typos_2024} & \textcolor{Red}{\faCircle} 0.04 & \textcolor{Red}{\faCircle} 0.00\\
 \hline
{\textbf{Jailbreak}} & Guardrails / Alignment \cite{stefano_rag_2024,anderson_is_my_data_2024,Nandagopal_2025,lu2024eraser} & \textcolor{Orange}{\faCircle} 0.38 & \textcolor{Orange}{\faCircle} 0.33 \\ 
\hline
\textbf{LLM Extraction} & Differential Privacy  \cite{wang_unique_2024} & \textcolor{Red}{\faCircle} 0.00 & \textcolor{Red}{\faCircle} 0.00 \\ \hline
 & Anonymization \cite{cohen_unleashing_2024,Ward_Harguess_2025} & \textcolor{Red}{\faCircle} 0.08 & \textcolor{Red}{\faCircle} 0.00\\
& Distance Metric Strategy \cite{zhou_trustworthiness_2024, cohen_unleashing_2024} & \textcolor{Red}{\faCircle} 0.08 & \textcolor{Red}{\faCircle} 0.00\\
\textbf{Membership} & Prompt Augmentation / Guardrails \cite{liu_mask-based_2024,anderson_is_my_data_2024,li_generating_2024} & \textcolor{Orange}{\faCircle} 0.42 & \textcolor{Orange}{\faCircle} 0.33\\
\textbf{Inference} & Retrieval Strategy (e.g. re-ranking) \cite{cohen_unleashing_2024,liu_mask-based_2024,li_generating_2024} & \textcolor{Orange}{\faCircle} 0.35 & \textcolor{Orange}{\faCircle} 0.33\\
& Prompt Rewriting / Rephrasing \cite{liu_mask-based_2024,li_generating_2024} & \textcolor{Orange}{\faCircle} 0.27 & \textcolor{Orange}{\faCircle} 0.33\\
& Text Summarization \cite{zhou_trustworthiness_2024} & \textcolor{Red}{\faCircle} 0.00 & \textcolor{Red}{\faCircle} 0.00\\ 
& Differential Privacy \cite{Feng_Zhang_Tian_Xu_Zhang_Zhu_Ding_Liu_2025_RAGLeak} & \textcolor{Red}{\faCircle} 0.00 & \textcolor{Red}{\faCircle} 0.00 \\ \hline
\multirow{3}{*}{\textbf{Prompt Extraction}}  
& Anonymization \cite{zhou_trustworthiness_2024,cohen_unleashing_2024,mehtasecure} & \textcolor{Red}{\faCircle} 0.12 & \textcolor{Red}{\faCircle} 0.00\\
\textbf{} 
& Prompt Augmentation / Guardrails \cite{agarwal_prompt_leakage_2024} & \textcolor{Red}{\faCircle} 0.12 & \textcolor{Green}{\faCircle} 1.00\\
& Prompt Rewriting / Rephrasing \cite{agarwal_prompt_leakage_2024} & \textcolor{Red}{\faCircle} 0.12 & \textcolor{Green}{\faCircle} 1.00\\
\hline
\multirow{3}{*}{\textbf{Prompt Injection}} & Guardrails \cite{stefano_rag_2024,Ward_Harguess_2025,Nandagopal_2025} & \textcolor{Red}{\faCircle} 0.12 & \textcolor{Red}{\faCircle} 0.00 \\
& Text Filtering \cite{pasquini_neural_2024} & \textcolor{Red}{\faCircle} 0.12 & \textcolor{Red}{\faCircle} 0.00 \\
& Grammar Checker \cite{pasquini_neural_2024} & \textcolor{Red}{\faCircle} 0.12 & \textcolor{Red}{\faCircle} 0.00\\
\hline
\end{tabular}
}
\label{tab:mapping}
\end{table}

\subsection{Mapping Mitigation Strategies}
\label{sec:mapping}
In Table \ref{tab:mapping}, we summarize the RAG privacy risk mitigations introduced above, and we map these strategies to the specific stage in the RAG pipeline where they might be implemented. Thus, each mitigation is directly associated with the type of leakage it can prevent, which is important for researchers and practitioners, not only for designing proper mitigation strategies, but also for better understanding the implications of placing privacy solutions at different points in the RAG process. Although we also map mitigations to adversarial attacks (Table \ref{tab:mapping}), we discuss in Section \ref{sec:process} that mitigating attacks and leakage can be viewed in tandem.

To perform such a mapping for each of the relevant papers proposing mitigations, we extracted the stage in the RAG pipeline where the risk occurs, i.e., where the mitigation is applied, and the primary threat to which it is linked (e.g., data leakage). It is important to note that we mapped a mitigation to a stage only if the paper explicitly applies or discusses it at that stage, and thus we did not infer or extrapolate possible applications from those stated in the works.

\begin{itemize}
    \itemsep 0em
    \item \textbf{HIGH}: dedicated section or implementation in the paper.
    \item \textbf{MID}: only discussed somewhere in the paper.
    \item \textbf{LOW}: simply mentioned as a potential mitigation.
\end{itemize}

Similar guidelines were also established for the \textit{maturity} score, which considered the reproducibility (R), cross-domain generalizability (G), and deployability (D) of the conducted experiments in a given work (one score assigned for each axis):

\begin{itemize}
    \itemsep 0em
    \item \textbf{HIGH}: code available, tests on more than one dataset, discussion of deployment costs/overhead.
    \item \textbf{LOW}: no code available, 0 or 1 datasets used, no mention of deployment considerations.
\end{itemize}

Finally, we aggregated the above categorizations into a relevance and maturity score for each of the mitigation categories, according to Equations \ref{eq:rel} and \ref{eq:mat}.

{\footnotesize
\begin{equation}
    \label{eq:rel}
    \textit{relevance} = (2 \times \# \text{HIGH}) + (1 \times \# \text{MID}) +(0.5 \times \# \text{LOW})
\end{equation}
\vspace{-5pt}
\begin{equation}
    \label{eq:mat}
    \textit{maturity} = \frac{1}{3}\sum\frac{\text{\#HIGH}_i}{\text{Total \# for Mitigation}}, i \in {\text{\{R, G, D\}}}
\end{equation}
}

The normalized scores for both relevance and maturity are presented in Table \ref{tab:mapping}, and are also binned for readability. We note that since the scores are normalized, a score of 0 does not mean the absence of relevance or maturity, but rather the lowest as compared to other mitigations, and vice versa.

\textbf{Worked Example.} From the five papers that proposed Differential Privacy as a mitigation for dataset leakage, only three implemented their solutions and thus were classified as HIGH relevance papers. The other two simply mentioned Differential Privacy as a possible mitigation and had therefore LOW relevance. The final Differential Privacy (dataset leakage) relevance score was computed as \begin{math} (2 \times \text{3 HIGH papers}) + (1 \times \text{0 MID papers}) +(0.5 \times \text{2 LOW papers}) = 7 \end{math}, resulting in a normalized value of \begin{math} 0.5 \end{math}. The Differential Privacy (dataset leakage) maturity score was computed as the average of the number of papers with high reproducibility (1 paper out of the total 5, so 20\%), high cross-domain
generalizability (only 3 papers out of the total 5, so 60\%), and high deployability which had a 0\% score, because none of the 5 papers took the cost or the latency of their method into consideration. The final maturity score resulted in \begin{math} \frac{(0.2 + 0.6 + 0)}{3} \approx 0.27 \end{math}.

\begin{table}[t]
\centering
\caption{An overview of datasets used in the evaluation setups of works addressing privacy in RAG.}
\scriptsize
\resizebox{0.99\linewidth}{!}{
\begin{tabular}{|p{0.02\linewidth}|p{0.33\linewidth}|p{0.68\linewidth}|}
\hline
\textbf{Type/Task} & \textbf{Dataset} & \textbf{Used In} \\ \hline
& Natural Questions \cite{kwiatkowski2019natural} &  \cite{cho_typos_2024,stefano_rag_2024,chaudhari_phantom_2024,zeng_mitigating_2024,liu_mask-based_2024,tan_glue_2024,li_generating_2024,cheng_trojanrag_2024,xue_badrag_2024,zhu_atm_2024,fang_enhancing_2024,altinisik_exploiting_2024,zhang_hijackrag_2024,shafran_machine_2024,zou_poisonedrag_2024,kuppa_manipulating_2024,nazary2025resource,Mo_Tang_Lin_Zhou_Li_2025,Jiao_Wang_Yang_2025,Zhang_Xin_Fang_Liu_Yi_Li_Liu_2025,Feng_Zhang_Tian_Xu_Zhang_Zhu_Ding_Liu_2025_RAGLeak}\\
 & MS-MARCO \cite{nguyen2016ms} & \cite{chaudhari_phantom_2024,liu_mask-based_2024,tan_glue_2024,cheng_trojanrag_2024,xue_badrag_2024,chen_black-box_2024,zhang_hijackrag_2024,shafran_machine_2024,zou_poisonedrag_2024,nazary2025resource,Mo_Tang_Lin_Zhou_Li_2025,Jiao_Wang_Yang_2025,Zhang_Xin_Fang_Liu_Yi_Li_Liu_2025} \\
 & HotpotQA \cite{yang2018hotpotqa} & \cite{chaudhari_phantom_2024,tan_glue_2024,cheng_trojanrag_2024,clop_backdoored_2024,roychowdhury_confusedpilot_2024,zhang_hijackrag_2024,zou_poisonedrag_2024,nazary2025resource,Mo_Tang_Lin_Zhou_Li_2025,Jiao_Wang_Yang_2025,Zhang_Xin_Fang_Liu_Yi_Li_Liu_2025} \\
& TriviaQA \cite{joshi2017triviaqa} & \cite{cho_typos_2024,zeng_mitigating_2024,li_generating_2024,zhu_atm_2024,fang_enhancing_2024} \\
\multicolumn{1}{|c|}{\textbf{General}} & WebQuestions \cite{berant2013semantic} & \cite{zeng_mitigating_2024,cheng_trojanrag_2024,zhu_atm_2024,fang_enhancing_2024} \\
\multicolumn{1}{|c|}{\textbf{Question}} & PopQA \cite{mallen2022not}& \cite{zhu_atm_2024}  \\
\multicolumn{1}{|c|}{\textbf{Answering}} & StrategyQA \cite{geva2021did} &\cite{chen_agentpoison_2024}\\
 & SQuAD \cite{rajpurkar2016squad} & \cite{cho_typos_2024,xue_badrag_2024} \\
 & Cosmos \cite{huang2019cosmos} & \cite{yu_textual_2024} \\
 & CuratedTrec \cite{baudivs2015modeling} & \cite{zeng_mitigating_2024} \\
 & RealtimeQA(-MC) \cite{kasai2023realtime} & \cite{xiang_certifiably_2024} \\
 & Quora & \cite{tan_glue_2024} \\ \hline
 & TextBook \cite{jin2021disease} & \cite{xian_vulnerability_2024} \\
 & StatPearls & \cite{xian_vulnerability_2024} \\
 & HealthCareMagic \cite{He_Tang_Zhang_Zhou_Su_2025,li2023chatdoctor} & \cite{zeng_good_2024,cohen_unleashing_2024,zeng_mitigating_2024,liu_mask-based_2024,anderson_is_my_data_2024,He_Liu_Hou_Jiang_Li_2025_PRESS,Feng_Zhang_Tian_Xu_Zhang_Zhu_Ding_Liu_2025_RAGLeak} \\
\multicolumn{1}{|c|}{\textbf{(Bio)medical}} & NFCorpus \cite{boteva2016full} & \cite{clop_backdoored_2024} \\
\multicolumn{1}{|c|}{\textbf{Datasets}} & MMLU-Med \cite{xiong_benchmarking_2024} & \cite{xian_vulnerability_2024,peng_data_2024} \\
 & MedQAUS \cite{xiong_benchmarking_2024} & \cite{xian_vulnerability_2024} \\
 & MedMCQA \cite{xiong_benchmarking_2024} & \cite{xian_vulnerability_2024,peng_data_2024} \\
 & PubMedQA \cite{xiong_benchmarking_2024,jin2019pubmedqa} & \cite{xian_vulnerability_2024, peng_data_2024} \\
 & BioASQ-Y/N \cite{xiong_benchmarking_2024,krithara2023bioasq} & \cite{xian_vulnerability_2024,bondarenko_llm_2024} \\ \hline
 & Pile \cite{gao2020pile} & \cite{kim2023propile} \\
 & FiQA \cite{maia201818} & \cite{tan_glue_2024} \\
\multicolumn{1}{|c|}{\textbf{General}} & Enron Emails \cite{enron_online} & \cite{zhou_trustworthiness_2024,zeng_good_2024,huang2023privacy_implications,li2023multi,zeng_mitigating_2024,anderson_is_my_data_2024,He_Liu_Hou_Jiang_Li_2025_PRESS} \\
\multicolumn{1}{|c|}{\textbf{NLP}} & WikiText \cite{merity2016pointer} & \cite{huang2023privacy_implications, zeng_mitigating_2024} \\
\multicolumn{1}{|c|}{\textbf{Datasets}} & WNUT 2017 \cite{derczynski2017results} & \cite{yu_textual_2024} \\
 & SST-2 \cite{socher-etal-2013-recursive} & \cite{cheng_trojanrag_2024} \\
 & AG News \cite{agnews} & \cite{cheng_trojanrag_2024} \\
 & MovieLens \cite{movielens} & \cite{nazary2025resource,Nazary_Deldjoo_Noia_2025}
 \\ \hline
\multicolumn{1}{|c|}{\textbf{Bias}} & BBQ \cite{parrish2021bbq} & \cite{cheng_trojanrag_2024} \\
\multicolumn{1}{|c|}{\textbf{and}} & AdvBench-V3 \cite{lu2024eraser} & \cite{cheng_trojanrag_2024} \\
\multicolumn{1}{|c|}{\textbf{Factuality}} & LLM Biographies \cite{min2023factscore} & \cite{xiang_certifiably_2024} \\ \hline
\end{tabular}
}
\label{tab:datasets}
\end{table}

\begin{table*}[t]
\centering
\caption{Evaluation metrics used in works addressing privacy in RAG.}
\scriptsize
\resizebox{0.97\textwidth}{!}{
\begin{tabular}{|p{0.133\linewidth}|p{0.64\linewidth}|p{0.45\linewidth}|}
\hline
\multicolumn{1}{|c|}{\textbf{Metric Name}} & \multicolumn{1}{c|}{\textbf{Description}} & \multicolumn{1}{c|}{\textbf{Used In}} \\ \hline
\multicolumn{3}{|c|}{\textbf{Retrieval Metrics}} \\ \hline
Accuracy & Metric for correctness of generated answers based on reference (e.g., top-k hit rate). & \cite{cho_typos_2024,yu_textual_2024,pasquini_neural_2024,li2023multi,liu_mask-based_2024,cheng_trojanrag_2024,xue_badrag_2024,chen_agentpoison_2024,jiao_can_2024,xiang_certifiably_2024,altinisik_exploiting_2024,ju_flooding_2024,Jiao_Wang_Yang_2025,Zhang_Xin_Fang_Liu_Yi_Li_Liu_2025,Feng_Zhang_Tian_Xu_Zhang_Zhu_Ding_Liu_2025_RAGLeak,bondarenko_llm_2024} \\
Precision / Recall & Metric for proportion and coverage of relevant contexts among the top-k retrieved ones. & \cite{liu_mask-based_2024,Chen_Tackman_Setälä_Poranen_Zhang_2025,He_Liu_Hou_Jiang_Li_2025_PRESS,cheng_trojanrag_2024,altinisik_exploiting_2024,zou_poisonedrag_2024,nazary2025resource,Mo_Tang_Lin_Zhou_Li_2025,Feng_Zhang_Tian_Xu_Zhang_Zhu_Ding_Liu_2025_RAGLeak} \\
F1-Score & Harmonic mean of precision and recall. &  \cite{li2024enhancing,liu_mask-based_2024,qi2024follow,cheng_trojanrag_2024,zhu_atm_2024,zhang_hijackrag_2024,zou_poisonedrag_2024,Mo_Tang_Lin_Zhou_Li_2025,Feng_Zhang_Tian_Xu_Zhang_Zhu_Ding_Liu_2025_RAGLeak}\\ \hline
\multicolumn{3}{|c|}{\textbf{Generation Metrics}} \\ \hline
ROUGE-N/-L & Metrics based on overlap of n-grams between generated and reference texts. & \cite{zeng_good_2024,yu_textual_2024,huang2023privacy_implications,zeng_mitigating_2024,qi2024follow,He_Tang_Zhang_Zhou_Su_2025,He_Liu_Hou_Jiang_Li_2025_PRESS,agarwal_prompt_leakage_2024,xue_badrag_2024,peng_data_2024,jiang_tc-ragturing-complete_2024} \\
BLEU-1/-4 & Precision-based metric that compares n-gram overlaps between generated and reference texts. & \cite{zeng_mitigating_2024,qi2024follow,He_Tang_Zhang_Zhou_Su_2025,He_Liu_Hou_Jiang_Li_2025_PRESS,jiang_tc-ragturing-complete_2024} \\
BERTScore & Similarity metric that measures the cosine similarity of BERT embeddings to compare generated and reference texts. & \cite{qi2024follow}  \\
LLM-as-a-Judge & A Large Language Model is used to evaluate the correctness, relevance, or quality of a generated response. & \cite{xue_badrag_2024,peng_data_2024,xiang_certifiably_2024,bondarenko_llm_2024} \\ \hline
\multicolumn{3}{|c|}{\textbf{Answer Metrics}} \\ \hline
Rejection Rate & Proportion of times the generator (model) refuses to answer. & \cite{zhou_trustworthiness_2024,xue_badrag_2024,zhang_agent_2024} \\
Benign Answers & Proportion or count of answers that are safe, correct, and contain no policy violations or harmful content. & \cite{stefano_rag_2024} \\
Malicious Answers & Proportion or count of answers that contain harmful, malicious, or disallowed content. &  \cite{stefano_rag_2024}\\
Ambiguous Answers & Proportion or count of answers that are unclear, vague, or could be interpreted in multiple ways. &  \cite{stefano_rag_2024}\\
Inconclusive Answers & Proportion or count of answers that do not provide a definitive statement. &  \cite{stefano_rag_2024}\\ \hline
\multicolumn{3}{|c|}{\textbf{Attack Metrics}} \\ \hline
Attack Success Rate & Percentage of attempts causing the system to reveal private content, or otherwise deviate from normal policy. & \cite{deng_pandora_2024,cho_typos_2024,wang_poisoned_2024,tan_glue_2024,xue_badrag_2024,clop_backdoored_2024,peng_data_2024,zhang_agent_2024,chen_agentpoison_2024,jiao_can_2024,zhang_hijackrag_2024,zou_poisonedrag_2024,Mo_Tang_Lin_Zhou_Li_2025,Jiao_Wang_Yang_2025,Zhang_Xin_Fang_Liu_Yi_Li_Liu_2025,bondarenko_llm_2024} \\
Retrieval Success Rate & Percentage of queries for which the system successfully retrieves the target documents, whether poisoned or not. & \cite{xian_vulnerability_2024,liu_mask-based_2024,tan_glue_2024,shafran_machine_2024,Zhang_Zhang_Lou_Wu_Wang_Chen_2025} \\
Retrieval Failure Rate & Percentage of queries for which the system fails to retrieve the target documents, whether poisoned or not. & \cite{chaudhari_phantom_2024} \\
Extraction Rate & Percentage of successful attempts of extracting the targeted data. & \cite{cohen_unleashing_2024} \\
Targeted information & Count of targeted information, such as poisoned documents or PII, that appear in the generated response. &  \cite{huang2023privacy_implications,li2023multi,zeng_mitigating_2024,zhang_human-imperceptible_2024,Zhang_Zhang_Lou_Wu_Wang_Chen_2025}\\ \hline
\multicolumn{3}{|c|}{\textbf{Other Metrics}} \\ \hline
Exact Match (Rate) & Evaluates if a prediction precisely matches the correct answer. &  \cite{li2024enhancing,cho_typos_2024,cheng_trojanrag_2024,zhu_atm_2024,jiang_tc-ragturing-complete_2024,kim2023propile}\\
Keyword Matching Rate & Recall rate between the reference and response based on ROUGE-L. & \cite{cheng_trojanrag_2024} \\
Mean Reciprocal Rank & Average reciprocal rank of the first relevant item in a ranked list of results. & \cite{hu_privacy-preserved_2024,chen_black-box_2024} \\
AUC ROC & Metric for evaluating the trade-off between true and false positive rates across thresholds. & \cite{liu_mask-based_2024,anderson_is_my_data_2024,li_generating_2024,zou_poisonedrag_2024,Feng_Zhang_Tian_Xu_Zhang_Zhu_Ding_Liu_2025_RAGLeak}\\ \hline
\end{tabular}
}
\label{tab:metrics}
\end{table*}

\subsection{Evaluating Privacy in RAG}
Important to studying RAG privacy risks and mitigations is the evaluation strategies undertaken in experimental setups. Recent works employ a wide variety of datasets and metrics to measure both the utility (RAG performance) and privacy (protection against risks) afforded by mitigation techniques. 

Widely used general question-answering datasets, such as MS-MARCO, HotpotQA, and TriviaQA, are also employed in most research addressing privacy in RAG systems. However, datasets for domain-specific tasks, particularly in (bio)medical, financial, and bias-related contexts, are also prevalent. For example, medical datasets such as TextBook, StatPearls, MedMCQA, MMLU-Med, and BioASQ are used to study privacy in settings involving sensitive health-related data. These datasets span a variety of domains and use cases, highlighting the broad applicability and relevance of RAG privacy research. 

Research conducting experiments on privacy in RAG systems utilizes a diverse set of evaluation metrics to empirically measure privacy protection. These can be categorized into five primary groups, providing a holistic view of how researchers analyze RAG performance under privacy-related conditions:
\begin{itemize}
    \item \textbf{Retrieval metrics}: assess the effectiveness of the retrieval component in isolating relevant information, especially in contexts where sensitive or adversarially injected data may be present. These metrics are critical in determining whether the system successfully retrieves harmful or private data, which is often the first step in privacy-adverse behaviors. Emphasis here is placed not only on the presence of correct documents but also on the balance between over-retrieval (which may include sensitive content) and under-retrieval (which could limit utility).
    \item \textbf{Generation metrics}: the focus shifts to the quality of the outputs generated based on the retrieved contexts. These methods are widely adapted from traditional natural language generation evaluation techniques but take on new relevance in privacy research, measuring to what extent privacy risk mitigations affect generation quality.
    \item \textbf{Answer metrics}: these metrics evaluate the content produced by the RAG system. This includes whether answers are benign, malicious, or ambiguous, or whether the model opts to refrain from answering altogether. These metrics are particularly useful for identifying indirect privacy risks, such as vague or misleading responses that may reflect underlying data exposure or misalignment with system policy. Thus, answer metrics view evaluation from a broader, ethical- and safety-focused lens.
    \item \textbf{Attack metrics}: measure the success of adversarial attempts to tamper with a RAG system. They can reveal the susceptibility of systems to prompt injection, poisoning, or targeted extraction of private data. They often differentiate between retrieval and generation failures, which is critical in tracing the propagation of an attack.
    \item \textbf{Other metrics}: these encompass auxiliary evaluation techniques, often borrowed from the machine learning and information retrieval disciplines. This might include precise matching and ranking-based metrics that quantify system accuracy and decision confidence, providing additional context to more targeted privacy evaluations.
\end{itemize}

A brief description of the diverse datasets and metrics for privacy evaluation in RAG are provided in Tables \ref{tab:datasets} and \ref{tab:metrics}.

\begin{figure*}[htbp]
    \centering
    \includegraphics[width=0.99\linewidth]{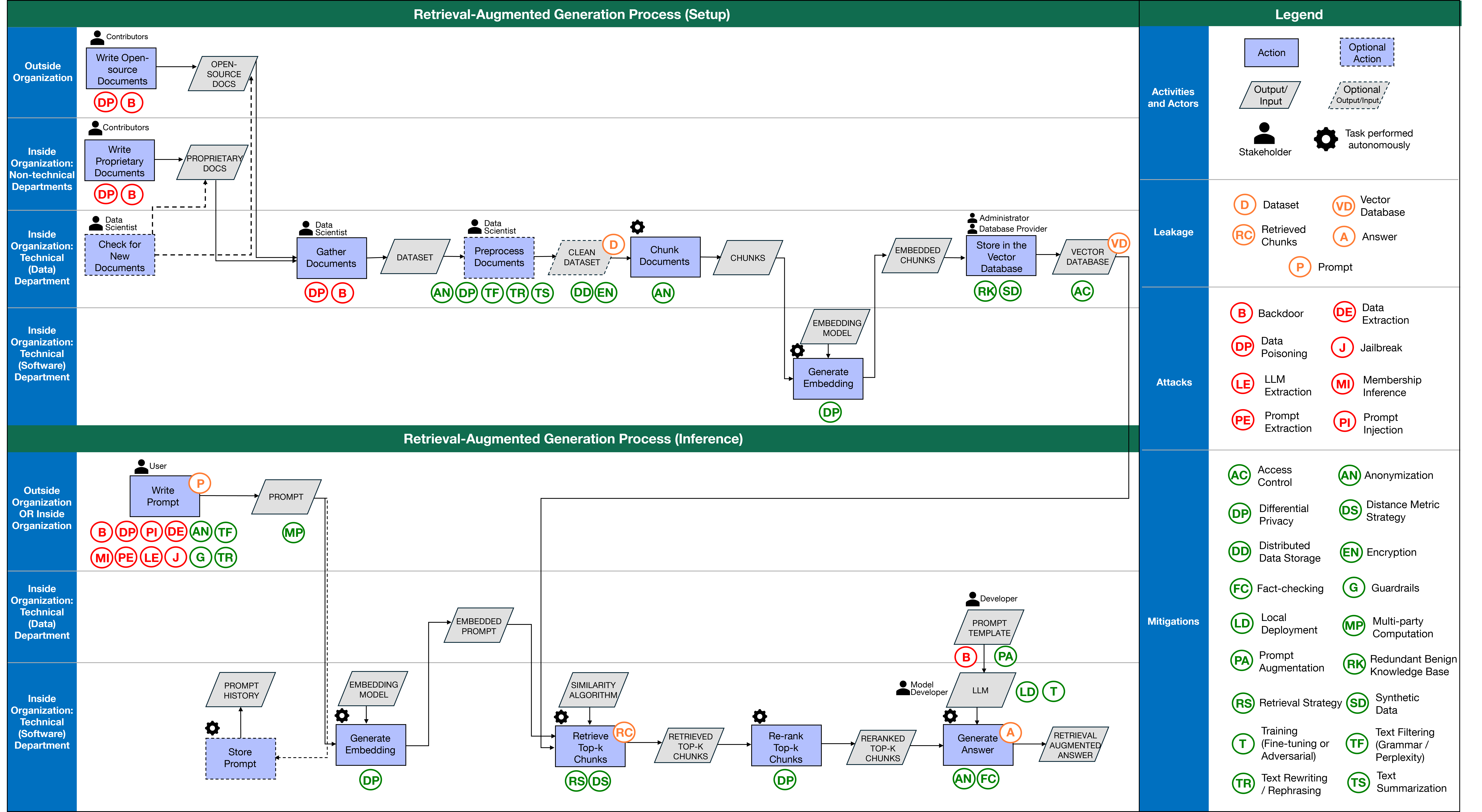}
    \caption{The RAG Privacy Process Diagram.}
    \label{fig:process}
\end{figure*}

\subsection{A Dynamic View of Privacy in RAG}
\label{sec:process}
Though the SLR sheds light on a number of privacy risks and proposed mitigations, the coverage of these findings hitherto represents a \say{static} view of the RAG privacy risk ecosystem. Specifically, the mapping presented in Table \ref{tab:mapping} is useful for directly associating proposed mitigation strategies and leakage types, yet there is no notion of how these mitigations might impact the remainder of the RAG pipeline; moreover, mitigating leakage should be better connected with the mitigation of associated adversarial attack vectors.

As such, we create a more dynamic view of the RAG privacy ecosystem, which we present in our second main artifact, the RAG Privacy Process Diagram (Figure \ref{fig:process}). This diagram is divided into \textit{RAG setup} and \textit{RAG inference} (i.e., runtime). In this ecosystem, there exists a separation of concerns between departments within an \say{organization}, or the entity responsible for the RAG system provision. Moreover, we distinguish between \textit{activities} carried out within this organization, and those taking place outside it. These activities are presented along with an indication of the \textit{actor} responsible for the task.

To bind the privacy risks (Figure \ref{fig:taxonomy}) and mitigations (Table \ref{tab:mapping}) to the complete RAG process, we indicate at which point risks \textit{originate}, and accordingly, \textit{where} mitigations serve to protect against such risks. With this perspective, one can visualize what further steps may be impacted by the realization of a privacy risk, or the introduction of a mitigation. This, above all, serves to contextualize decisions made in privacy risk mitigation plans, showing that such decisions cannot be made in isolation from a particular threat actor or RAG process activity. We note that the diagram models a \say{naive} (simple) RAG pipeline; advanced RAG setups were out-scoped.

\section{Discussion}
In this section, we reflect on the findings of our survey on privacy in RAG, critically assess the current state of privacy mitigations, and discuss implications for future research.

\textbf{What is new with privacy in RAG?}
Our literature review shows a clear increase in research attention paid to risks associated with RAG systems, with just three relevant papers in 2023 to 51 papers in 2024. This not only points to the growing importance of the topic, but also to the diversity and complexities within the research field of privacy in RAG.

A survey of the privacy risks in RAG systems reveals three important factors that exacerbate the threat of known adversarial attacks, as well as create complexities in the elicitation of novel RAG privacy risks. Firstly, the usefulness of RAG in bringing life to typically \say{static} LLMs is certainly proven, yet it is a double-edged sword in the way that \say{live} data presents an especially vulnerable point that LLMs alone do not exhibit, i.e., exposed private data. As a second and related point, the ultimate goal of RAG systems is to make such data more accessible by allowing users, whether internal or external to organizations, to interface with the data. This naturally creates new risks of improper data disclosure without proper measures. Such measures are two-fold in the sense that external safeguards must be in place, such as access control, alongside system internal mitigations, such as anonymization. 

This gives way to the third aspect of RAG systems, leading to novelties in the study of privacy risks, which relates directly to the \say{system} nature of RAG itself. As opposed to studying privacy risks in LLMs, for example, RAG systems exhibit many points of potential failure or compromise, and these points take various forms, e.g., raw text data, vectorized document chunks, or LLM-generated answers. This typical RAG pipeline is incredibly dynamic, involving multiple stakeholders and individual technologies, presenting not only a wide attack surface, but also a more complex ecosystem for mitigations to be implemented. Thus, whereas many of the potential privacy threats to RAG systems may be similar to known risks from a technical point of view, the landscape of RAG systems opens the door to new avenues for attack realization and mitigation.

\textbf{It's all about data leakage, but leakage can mean many things!}
Upon our initial reading of the selected primary sources, we learned that many of the perceived privacy risks relating to RAG systems revolve around the form of \textit{leakage}. As displayed in our Taxonomy of RAG Privacy Risks (Figure \ref{fig:taxonomy}), this is only one side of the story, and privacy risks are often studied from the angle of \textit{adversarial manipulation}, or designed attacks. This is exemplified by Table \ref{tab:system}, which shows a near parity between papers focusing on the leakage or adversarial aspects. Together, \textit{leakage} and \textit{attacks} form the two-sided coin of RAG privacy risks, with the former focusing on \textit{what} is exploited, while the latter focuses on the \textit{how}.

The story runs deeper, however, and we learn that data leakage cannot be viewed as a single phenomenon, but rather a spectrum of possible privacy vulnerabilities. This largely roots itself in the fact that data passes through many \say{checkpoints} in the RAG pipeline, and at each of these, it may be re-encoded, chunked, or harmonized with the help of LLMs. These distinct stages make the investigation of RAG privacy risks dynamic, and a useful way to reason about such risks is to intertwine the RAG stages with points where leakage may originate. Thus, we distinguish leakage that can be pinpointed to the \textit{system} (data leakage) or the \textit{user} (prompt leakage).

As a final piece to the story, we find that viewing privacy risks and their mitigations in isolation (Table \ref{tab:mapping}) is useful but not completely satisfactory. To contextualize such a mapping further, we create a more dynamic view of the privacy risks in RAG, in the form of the RAG Privacy Process Diagram (Figure \ref{fig:process}). In this, we propose two primary improvements to the study of RAG privacy risks: (1) the perspective of risks and mitigations as part of a larger ecosystem, where either the fruition of a risk or the implementation of a mitigation carries effects downstream, and (2) a more illustrative picture of where mitigations can be implemented in the pipeline, and for which risks (attacks) there may not currently be sufficient protections. A prime example of the latter point surfaces at the user interface of RAG systems, which represents the stage at which many RAG privacy risks are realized, yet where more proposals for novel mitigations can be made. Thus, we hope that the process diagram becomes a living artifact, where future updates may serve to track the progress made, as well as the new risks arising, regarding privacy in RAG.

\textbf{Mitigating leakage: viewing the RAG privacy process in action.}
As previously introduced and in light of Figure \ref{fig:process}, it becomes interesting to explore the impact of mitigation techniques on the overall functioning of the RAG system. The pipeline nature of RAG, in which data exists in many forms (raw text, embedding vectors, text chunks), introduces complexities for the implementation of mitigations, in that the efficacy and trade-off resulting from the mitigation are directly affected by the form of inputs as well as the nature of the following states. As an example, performing anonymization directly on the raw text data might be very effective for removing sensitive information, but may degrade the quality of the resulting embeddings, text chunk retrieval, and answer generation. In contrast, implementing mitigations further downstream may serve to preserve utility more effectively, but at the cost of late or even post-hoc privatization. 

What is uncertain from current research, however, is the effectiveness of mitigations \textit{in sequence}. While it may seem wise to implement protections at various points along the RAG process, we found no evidence of such experimentation. Beyond feasibility, we envision that determining effective ensembles of mitigations would require meticulous research. This comes in addition to fundamental research on the strengths and limitations of mitigations at different stages, e.g., database anonymization versus generated answer rewriting.

As such, we hope that with the guidance of the RAG Privacy Process Diagram, future studies can focus on investigating \textit{mitigations in context}, giving credence to the merits of proposed mitigations, while also uncovering their potential limitations.

\textbf{The state of current mitigations.}
In order to investigate deeper the main focus of the works we survey, we also systematize the balance in the current literature between works focusing on adversarial attacks affecting privacy, and those that propose mitigations to combat these (Table \ref{tab:system}). We find a relatively significant skew towards \say{attack papers} (56\%), whereas papers focusing specifically on privacy mitigations comprise only 36\% of the reviewed papers (with 8\% in neither category). While it is also the case that attack papers often test mitigations, they rarely are the primary focus, suggesting the need for an uptick in privacy mitigation research for RAG.

We also quantify the \textit{relevance} and \textit{maturity} of proposed mitigations (Table \ref{tab:mapping}). These scores, while imperfect approximations of the current state of privacy mitigations in RAG, provide an overall sense of the relative attention a certain mitigation strategy has received (\textit{relevance}), as well as how often they are practically tested (\textit{maturity}). A mitigation with a high relevance but low maturity may imply that many works have proposed the mitigation, but fewer have tested it. On the other hand, higher maturity than relevance would suggest that, although the raw quantity of mentions may be less, such a method may be more mature since it has been relatively more often evaluated. With this, one can see that mitigations for certain risks, such as dataset leakage and data poisoning, are generally more mature than those of LLM extraction, backdoor, or prompt injection attacks, for example. This assessment, therefore, sheds light on the potentially under-researched mitigation areas for RAG privacy risks.

As can be extrapolated from Tables \ref{tab:system} and \ref{tab:mapping}, the state of current privacy risk mitigations in RAG leans on the immature end, made evident by the generally low \textit{relevance} and \textit{maturity} scores, as well as the relatively low amount of works that publish code to reproduce either attacks or mitigations (roughly 40\%). This would suggest that reasoning about privacy risks and mitigations in RAG, even in the research sphere, still exists in the \say{ideation} phase, with growing numbers of actionable artifacts being published. This is also made clear by the low amount of papers discussing practical considerations of mitigation adoption (24\%), such as computational speed or resources required to run such proposed mitigations.

With our systematization and mitigation assessments, we strive for these scores to form a foundation for quantifying the current state of mitigation strategies for RAG privacy, and moreover, to provide a practical sense of \textit{feasibility} for researchers and practitioners going forward. In this light, we synthesize our findings into a set of key focus areas for future research on the mitigation of privacy risks in RAG, focusing on the recurring challenges we observed from the literature:

\begin{itemize}
    \itemsep 0em
    \item \textbf{More practical considerations}: as previously noted, we found that only about a quarter of papers implementing mitigations also provided details on computational costs and other practical deployment considerations. We see this is an important area to improve to increase the practical applicability of privacy risk mitigations in RAG.
    \item \textbf{Real-world testing}: we found very few cases of exposing privacy risks or testing mitigations on live RAG systems, implying a potential disconnect between current research and real-world privacy concerns in RAG. Furthermore, our review of grey literature suggested that practical privacy solutions currently do not match the intricacy of those proposed in the literature, which likewise strengthens the need for practicality in RAG privacy research.
    \item \textbf{All RAG is not the same}: we found little evidence of considering various RAG setups, leaving the question unanswered how privacy is affected (for better or for worse) by more complex RAG architectures, including those implementing agentic systems.
    \item \textbf{Towards RAG-specific privacy evaluation}: particularly in the use of datasets for RAG privacy evaluation, we perceive that many of the leveraged datasets (Table \ref{tab:datasets}) are \say{reused} from other purposes, and there is a lack of privacy-specific benchmark datasets for RAG use cases. To an extent, this also applies to evaluation metrics (Table \ref{tab:metrics}), which often relate to either pure RAG evaluation or more security-specific measurements (e.g., rejection rate).
\end{itemize}

\section{Related Work}
\textbf{Surveying RAG.}
Previous works survey various aspects of RAG systems, such as general applications \cite{zhao_retrieval-augmented_2024, hu_rag_2024}, architectures and optimization strategies \cite{gao2024retrievalaugmentedgenerationlargelanguage}, and evaluation strategies \cite{li_matching_2024, yu_evaluation_2024}. Other works investigate more specific aspects of RAG, such as trustworthiness \cite{fan_survey_2024,zhou_trustworthiness_2024} or RAG with multimodal data \cite{zhao-etal-2023-retrieving}. These works, however, make no or very tangential connections to the topic of privacy in RAG.

In exploring privacy in RAG, Zeng et al. \cite{zeng_good_2024} consolidate a number of adversarial attack types and propose defense strategies to mitigate them. 
While this survey offers one of the most detailed treatments of privacy in RAG systems in the current literature, it is limited in its scope of literature coverage, and it does not follow a formal methodology. 
In contrast, our work is guided by an Systematic Literature Review, analyzing how specific architectural and procedural elements in RAG pipelines lead to privacy risks, and we systematize a larger body of works on privacy aspects in RAG.

\textbf{Privacy in LLMs.}
Beyond RAG systems, multiple other works consider the privacy implications of LLMs. In their survey, Wang et al. \cite{wang_unique_2024} also relate privacy risks in LLMs to those in RAG, organizing security and privacy threats of LLMs across five stages of their lifecycle: pre-training, fine-tuning, deployment, LLM agents, and RAG. Their analysis of privacy in RAG remains high-level, with RAG only discussed as part of a broader LLM lifecycle. However, the view of privacy risks as part of a \textit{process} motivated the broader contextualization of RAG privacy risks in our work. In addition to other surveys exploring privacy in LLMs \cite{smith2023identifying,yan2024protecting}, recent works view privacy risks from the standpoint of (generative) AI \cite{10478883}, including in the context of the AI lifecycle \cite{10155147}, practical perspectives \cite{294558,klymenko2025we}, and general user perceptions \cite{10.5555/3632186.3632218}. 

\section{Conclusion}
We systematize the extant literature investigating privacy risks and proposed mitigations in RAG. We find that privacy risks in RAG systems can be categorized into two primary categories, leakage and adversarial manipulation, which can be mapped to a variety of innovative mitigation techniques. In the evaluation of privacy in RAG systems, a wide variety of datasets and metrics have been utilized, pointing to a wealth of potential evaluation strategies, but also to a general lack of unification. To augment our Taxonomy of RAG Privacy Risks, we contextualize risks and mitigations in a RAG Privacy Process Diagram, which acknowledges the dynamic nature of RAG and the confluence of risks, actors, and potential mitigations within this pipeline. Together, our findings not only illuminate the ecosystem of privacy risks in the context of RAG, but also map the current progress of mitigation efforts, providing a foundation for future studies at the intersection of privacy, RAG, and responsible AI.

\textbf{Limitations.}
We acknowledge a number of threats to validity, particularly concerning the conduction of the SLR. Firstly, the literature exclusion and filtering process was carried out solely by the primary researcher, introducing the possibility for researcher bias and subjectivity. Likewise, the full literature reading was carried out by this researcher. To mitigate this bias, weekly meetings were held with the larger research team over the course of the study, in order to validate the data extracted, as well as to make decisions during the systematization process (e.g., how to group mitigations). We also performed double coding of the structured literature analysis (data extraction) by two additional researchers on the team. In the literature review, we also considered more established non-published / non-peer-reviewed preprints, which could potentially have affected the validity of the final presented artifacts.

We also caution that our survey and systematization efforts, including the resulting artifacts, were studied in the scope of \say{simpler} (naive) RAG pipelines, as depicted in Figure \ref{fig:RAG}. As such, we did not generally account for the intricacies, and potential exacerbating factors to privacy risks, of more complex or advanced RAG architectures. We leave this as future work to build on our study results.

\textbf{Outlook and Future Work.}
Our work systematizes a continually expanding field, and we envision that our findings may serve to ground future research, particularly from a clearer definition of \textit{what} is being mitigated (Figure \ref{fig:taxonomy} and Table \ref{tab:mapping}) and \textit{how} this fits within the larger RAG pipeline (Figure \ref{fig:process}). From a practical perspective, we bring greater awareness to the potential risks of hosting RAG systems, as well as a contextualization of state-of-the-art mitigation techniques and their current research attention and maturity. We see three important points for future research: (1) validation of our proposed artifacts, specifically the RAG Privacy Process Diagram, (2) extensive experiments and feasibility studies involving privacy mitigations at various points in the RAG pipeline, and (3) studying user perceptions of RAG privacy risks, as well as tolerance for trade-offs introduced by mitigation techniques.

\bibliographystyle{IEEEtran}
\bibliography{bibliography}
\balance

\appendix
\section{Supplemental Materials}
All supplemental materials, including a full list of literature sources, our coded literature analysis, and complete table for the mitigation scoring, can be found in our public GitHub repository: \url{https://github.com/sebischair/SoK-RAG-Privacy}

\end{document}